\documentclass[12pt]{article}
\usepackage[centertags]{amsmath}
\usepackage{amsfonts}
\usepackage{amssymb}
\usepackage{amsthm}
\usepackage{newlfont}

\numberwithin{equation}{section}

\newtheorem{theorem}{Theorem}

\newtheorem{definition}{Definition}

\newtheorem{corollary}{Corollary}

\begin{document}

\title{\textbf{Generalized geometric commutator theory and quantum geometric bracket and its uses}}

\author{Gen WANG\thanks{School of Mathematical Sciences, Xiamen University,
     Xiamen, 361005, P.R.China. email:  wanggen@zjnu.edu.cn
}}
\date{}

\maketitle

\begin{abstract}
Inspired by the geometric bracket for the generalized covariant Hamilton system, we abstractly define a generalized geometric commutator $$\left[ a,b \right]={{\left[ a,b \right]}_{cr}}+G\left(s,a,b \right)$$ formally equipped with geomutator $G\left(s, a,b \right)=a{{\left[ s,b \right]}_{cr}}-b{{\left[ s,a \right]}_{cr}}$ defined in terms of structural function $s$ related to the structure of spacetime or manifolds itself for revising the classical representation ${{\left[ a,b \right]}_{cr}}=ab-ba$ for any elements $a$ and $b$ of any algebra.

Then we use the generalized geometric commutator to define quantum covariant Poisson bracket that is related to the quantum geometric bracket defined by geomutator as a generalization of quantum Poisson bracket.  The covariant dynamics includes the generalized Heisenberg equation as a natural extension of Heisenberg equation and G-dynamics based on the quantum geometric bracket, meanwhile, the geometric canonical commutation relation is induced.  As an application, we reconsider the canonical commutation relation and the quantization of field to be more complete.
\end{abstract}
\newpage
\tableofcontents

\section{Introduction}
\label{sec:intro}
Our goal in this paper is to propose the following result.
\begin{definition}[QCPB]
  The quantum covariant Poisson bracket in terms of quantum operator $\hat{f},~\hat{g}$ is generally defined by
 \[\left[ \hat{f},\hat{g} \right]={{\left[ \hat{f},\hat{g} \right]}_{QPB}}+G\left(s, \hat{f},\hat{g} \right)\] where $G\left( s, \hat{f},\hat{g} \right)=\hat{f}{{\left[ s,\hat{g} \right]}_{QPB}}-\hat{g}{{\left[ s,\hat{f} \right]}_{QPB}}$ is called quantum geometric bracket satisfying $G\left( s, \hat{f},\hat{g} \right)=-G\left( s, \hat{g},\hat{f} \right)$.
\end{definition}
Our method of new quantum bracket is inspired by the ideas of the GSPB \footnote{GSPB: Generalized Structural Poisson Bracket; QCPB: Quantum Covariant Poisson Bracket}\cite{22} which can copy the same pattern as Dirac \cite{1,2} did before.
This GSPB for the generalized covariant Hamilton system is one of the main things that inspired me to deeply develop a new quantum bracket that is complete and compatible.  The quantum covariant Poisson bracket as a new tool can mathematically help us better understand and analyze how the quantum mechanics essentially works.

That is deep reason at the same time realizing that we're probably will have much time to reach an incredible field out there and we're trying to be optimistic to seek more secrets about quantum world.  Quantum geometric bracket defined by
geomutator is such a beautiful thing ever existed in the universe. New theory has begun to paint a stunning picture of how the quantum secrets might unfold.

This quantum covariant Poisson bracket takes us on a part of journey to the truth of quantum mechanics,  to discover what the fate of quantum universe may ultimately be.  This is a picture of the covariant quantum mechanics as painted by quantum geometric bracket---a picture that will surely evolve over time as we dig for more clues to how our story will unfold. Much of the content of it is very recent-and hence new puzzle pieces are still waiting to be found in quantum realm.  Indeed, this overhead view of quantum geometric bracket gives a profound perspective-that the quantum world is closely connected with the geometric structure.

\subsection{Dirac's method for canonical quantization}
Dirac's method for canonical quantization, when it was firstly developed, quantum physics dealt only with the quantization of the motion of particles. By conventionally stated, the original form of particle quantum mechanics is denoted as first quantization, while quantum field theory is formulated in the language of second quantization.

The following exposition is based on Dirac's treatise on quantum mechanics for canonical quantization.  According to Dirac's method \cite{1,2}, Poisson's bracket $\left\{ \cdot, \cdot\right\}_{PB}$ is connected with commutator $ \frac{1}{\sqrt{-1}\hbar }\left[ \cdot, \cdot\right]_{QPB}$:
\[{{\left[ \hat{f},\hat{g} \right]}_{QPB}}=\sqrt{-1}\hbar \widehat{{{\left\{ f,g \right\}}_{PB}}}\]
where $\hat{\cdot }:f\mapsto \hat{f}$ is mapping real valued functions $f\left( {{q}_{i}},{{p}_{j}} \right)$ in phase space to real operators $\hat{f}$ in Hilbert space.

Recall that the relation of classical and quantum brackets.
Dirac's book \cite{1,2} details his popular rule of supplanting Poisson brackets by commutators:
\begin{equation}\label{eq2}
  \{f,g\}_{PB} \longmapsto \tfrac{1}{\sqrt{-1} \hbar} [\hat{f},\hat{g}]_{QPB}
\end{equation}
One might interpret this proposal as saying that we should seek a quantization map $Q$ mapping a function $f$ on the classical phase space to an operator ${\displaystyle Q_{f}}$ on the quantum Hilbert space such that \cite{1,2,8,9}
$${\displaystyle Q_{\{f,g\}_{PB}}={\frac {1}{\sqrt{-1}\hbar }}[Q_{f},Q_{g}]_{QPB}}$$
It is now known that there is no reasonable such quantization map satisfying the above identity exactly for all functions $f$ and $g$.

In quantum mechanics, quantum Poisson brackets $\left[ \hat{f},\hat{g} \right]_{QPB}$ is the commutator of two operators $\hat{f},~\hat{g}$  defined as $\left[ \hat{f},\hat{g} \right]_{QPB}=\hat{f}\hat{g}-\hat{g}\hat{f}$. This relation also holds for classical mechanics, the classical limit of the above, given the correspondence between Poisson brackets and commutators, \cite{3,4,5,6,7}
$$ {\displaystyle [\hat{f},\hat{H}]_{QPB}\quad \longleftrightarrow \quad \sqrt{-1}\hbar \{f,H\}}_{PB}$$
In classical mechanics, for a function $f$ with no explicit time dependence,
$df/dt={\displaystyle \{f,H\}_{PB}}$.
This approach also has a more direct similarity to classical physics: by simply replacing the commutator above by the Poisson bracket, the Heisenberg equation reduces to an equation in Hamiltonian mechanics.
The Dirac bracket is a generalization of the Poisson bracket developed by Paul Dirac to treat classical systems with second class constraints in Hamiltonian mechanics, and to thus allow them to undergo canonical quantization. It is an important part of Dirac's development of Hamiltonian mechanics to elegantly handle more general Lagrangians.

The possible results of a measurement are the eigenvalues of the operator representing the observable which explains the choice of Hermitian operators, for which all the eigenvalues are real. The probability distribution of an observable in a given state can be found by computing the spectral decomposition of the corresponding operator. Heisenberg's uncertainty principle is represented by the statement that the operators corresponding to certain observables do not commute.

In quantum mechanics in the Heisenberg picture the state vector, $\left| \psi  \right\rangle $  does not change with time, while an observable $\hat{A}$ is a physical quantity that can be measured which satisfies the Heisenberg equation of motion \cite{1,2,3,4,5,6,7}.
$$\frac{d\hat{A}}{dt} = {\sqrt{-1}\over \hbar } [ \hat{H}, \hat{A} ]_{QPB}  + \frac{\partial \hat{A}}{\partial t}~$$
Usually, observables are given by self-adjoint operators that include position and momentum. In systems governed by classical mechanics, it is a real-valued function on the set of all possible system states. In quantum physics, it is an operator, where the property of the quantum state can be determined by some sequence of operations. For example, these operations might involve submitting the system to various electromagnetic fields and eventually reading a value.
Dynamical variables $\hat{A}$ such as position, translational linear momentum, orbital angular momentum, spin, and total angular momentum are each associated with a Hermitian operator ${\hat {A}}$ that acts on the state of the quantum system.

Commutator relations may look different than in the Schr\"{o}dinger picture, because of the time dependence of operators. For example, consider the operators $x(t_{1}), x(t_{2}), {{\hat{p}}}^{{\left( cl \right)}}(t_{1})$ and ${{\hat{p}}}^{{\left( cl \right)}}(t_{2})$. The time evolution of those operators depends on the Hamiltonian of the system. Considering the one-dimensional harmonic oscillator, \cite{4,5,6,7}
\begin{equation}\label{eq11}
   \hat{H}^{\left( cl \right)}={\frac  {{{\hat{p}}}^{{\left( cl \right)2}}}{2m}}+{\frac  {m\omega ^{{2}}x^{{2}}}{2}}
\end{equation}
the evolution of the position and momentum operators is given by:
\begin{equation}\label{eq10}
  {d \over dt}x(t)={\sqrt{-1} \over \hbar }[ \hat{H}^{\left( cl \right)},x(t)]_{QPB}={\frac  {{{\hat{p}}}^{{\left( cl \right)}}}{m}}
\end{equation}\[{d \over dt}{{\hat{p}}}^{{\left( cl \right)}}(t)={\sqrt{-1}\over \hbar }[ \hat{H}^{\left( cl \right)},{{\hat{p}}}^{{\left( cl \right)}}(t)]_{QPB}=-m\omega ^{{2}}x\]
As for the application of the QPB or the commutation, there are some many fields including the physics and mathematics, and so on. In particular, the SUSY, but there are still problems unsolved.  This is why we reconsider the QPB or the commutation.

The canonical commutation relation \cite{1,2,3,4,5,6,7,8,9,10,11,12,13,14,15} is the fundamental relation between canonical conjugate quantities. For example, a set of equal-time commutation relations are introduced as
\begin{equation}\label{eq1}
  \left[ {{{x}_{i}}},{{\hat{{{p}_{j}}}}^{\left( cl \right)}} \right]_{QPB}=\sqrt{-1}\hbar {{\delta }_{ij}}
\end{equation}
between the position operator $\hat{{{x}_{i}}}$ and classical momentum operator ${{\hat{{{p}_{j}}}}^{\left( cl \right)}}$ in the $\hat{{{x}_{i}}}$ direction of a point particle in three dimension, $\hbar$ is the reduced Planck's constant, where $\delta _{ij}$ is the Kronecker delta.
This observation led Dirac \cite{1,2} to propose that the quantum counterparts ${\hat {f}},{\hat {g}}$ of classical observables $f, g$ satisfy \cite{1,2,3,4}
\begin{equation}\label{eq8}
  [\hat f,\hat g]_{QPB}=\sqrt{-1} \hbar\widehat{\{f,g\}}_{PB}
\end{equation}
In the formalism of quantum mechanics, the state of a system at a given time is described by a complex wave function, also referred to as state vector in a complex vector space.
Consider a dynamical variable $\hat{f}$ corresponding to a fixed linear operator in the Schr\"{o}dinger picture.   Heisenberg equation of motion can be written \cite{1,2,3,4,5,6,7,8,9,10,11,12,13,14,15}
\begin{equation}\label{eq9}
  \frac{d\hat{f}}{dt}=\frac{1}{\sqrt{-1}\hbar }{{\left[ \hat{f},\hat{H} \right]}_{QPB}}
\end{equation}
where $\hat{H}$ is the Hamiltonian and $\left[ \cdot ,~\cdot  \right]_{QPB}$ denotes the commutator of two operators.
For the time-independent operators $\hat {A}$, we have $\partial \hat {A}/\partial t = 0$, therefore,  the Heisenberg equation above reduces to:
$\sqrt{-1}\hbar {d\hat {A} \over dt}=[\hat {A},\hat {H}]_{QPB}$.
Above equation shows how the dynamical variables of the system evolve in the Heisenberg picture. It is denoted the Heisenberg equation of motion. Note that the time-varying dynamical variables in the Heisenberg picture are usually called Heisenberg dynamical variables to distinguish them from Schr\"{o}dinger dynamical variables, which do not evolve in time. Canonical Hamilton equations of motion is represented by \cite{1,2,3,4,5,6,7,8,9,10,11,12,13,14,15}
\begin{align}
  & \frac{d}{dt}{{{x}}_{j}}  =\frac{1}{\sqrt{-1}\hbar }{{\left[ {{{x}}_{j }},\hat{H} \right]}_{QPB}} \notag\\
 & \frac{d}{dt}{{\hat{p}}_{i }}   =\frac{1}{\sqrt{-1}\hbar }{{\left[ {{\hat{p}}_{i}},\hat{H} \right]}_{QPB}} \notag
\end{align}
Based on the Poisson bracket and Lie bracket have ingenious similarity, the formal is for classic mechanics, while the latter is used for the  quantum mechanics.
Therefore, this paper will follow the step of Dirac's method to complete the some rules of quantum mechanics.
The goal of this paper is to define a completely new bracket which is called quantum covariant Poisson bracket (QCPB) to analyze the incomplete or missing parts of quantum mechanics.

\subsection{Quantization of Field}
As we already know, quantization procedure is to generally convert classical fields into operators acting on quantum states of the field theory. The first method to be developed for quantization of field theories was canonical quantization. While this is extremely easy to implement on sufficiently simple theories, there are many situations where other methods of quantization yield more efficient procedures for computing quantum amplitudes.
Canonical quantization of a field theory is analogous to the construction of quantum mechanics from classical mechanics. The classical field is treated as a dynamical variable called the canonical coordinate, and its time-derivative is the canonical momentum. One introduces a commutation relation between these which is exactly the same as the commutation relation between a particle's position and momentum in quantum mechanics. This procedure can be applied to the quantization of any field theory: whether of fermions or bosons, and with any internal symmetry.

The canonical quantization is a standard procedure for quantizing a classical theory, while attempting to preserve the formal structure, such as symmetries.
The canonical arises from the Hamiltonian approach to classical mechanics, where a system's dynamics is generated via canonical Poisson brackets, a structure which is only partially preserved in canonical quantization.

\subsubsection{Canonical quantization of Bose field}
The quantum of a Bose field, the spin is integer or zero, obeying the Bose-Einstein statistical method, known as bosons, the canonical coordinate ${{\varphi }_{i}}$ and the canonical conjugate momentum ${{p}_{i}}$ of the field are regarded as the operator of Hilbert space, and the corresponding quantum commutation relation in discrete field form is \cite{16,17}
$$\left[ {{\varphi }_{i}},{{p}_{j}} \right]_{QPB}=\sqrt{-1}\hbar {{\delta }_{ij}}$$
$$\left[ {{\varphi }_{i}},~~{{\varphi }_{j}} \right]_{QPB}=\left[ {{p}_{i}},{{p}_{j}} \right]_{QPB}=0$$
Canonical equations of motion for corresponding quantum fields are shown as
$$\overset{\centerdot }{\mathop{\varphi }}\,=\frac{\sqrt{-1}}{\hbar }\left[ \hat{H},\varphi  \right]_{QPB},~~\overset{\centerdot }{\mathop{\pi }}\,=\frac{\sqrt{-1}}{\hbar }\left[ \hat{H},\pi  \right]_{QPB}$$The quantization of the field depends entirely on the commutators.

\subsubsection{Canonical quantization of Fermi field}
A fermion is a particle that follows Fermi-Dirac statistics. These particles obey the Pauli exclusion principle. Fermions include all quarks and leptons, as well as all composite particles made of an odd number of these, such as all baryons and many atoms and nuclei. Fermions differ from bosons, which obey Bose-Einstein statistics.
Suppose that $\Psi \left( t,x \right)$ is the field function of the Fermi field, and the canonical coordinates and canonical conjugate momenta of the field are $\Psi \left( t,x \right)$, $\pi \left( t,x \right)$ respectively, the canonical motion equation of quantum field is \cite{16,17}
\begin{align}
  & \overset{\centerdot }{\mathop{\Psi }}\,\left( t,x \right)=\frac{\sqrt{-1}}{\hbar }\left[ \hat{H}\left( t \right),\Psi \left( t,x \right) \right]_{QPB} \notag\\
 & \overset{\centerdot }{\mathop{\pi }}\,\left( t,x \right)=\frac{\sqrt{-1}}{\hbar }\left[ \hat{H}\left( t \right),\pi \left( t,x \right) \right]_{QPB} \notag
\end{align}

\section{Generalized structural Poisson bracket and geobracket}
In this section, we will briefly review the entire theoretical framework of generalized covariant Hamiltonian system defined by the generalized structural Poisson bracket totally based on the paper \cite{22} as a revision of generalized Poisson bracket.

To begin with the generalized Poisson bracket \cite{18,19,20} which
is defined as the bilinear operation
\[{{\left\{ f,g \right\}}_{GPB}}={{\nabla }^{T}}fJ\nabla g={{J}_{ij}}\frac{\partial f}{\partial {{x}_{i}}}\frac{\partial g}{\partial {{x}_{j}}}\] where structural matrix $J$ satisfies antisymmetric ${{J}_{ij}}={{\left\{ {{x}_{i}},{{x}_{j}} \right\}}_{GPB}}=-{{J}_{ji}}$. The
generalized Hamiltonian system (GHS) is defined as \cite{18,22}
\[{\dot{x}}=\frac{dx}{dt}=J\left( x \right)\nabla H\left( x \right),~~~x\in {{\mathbb{R}}^{n}}\]
where
$J\left( x \right)$ is structure matrix, $\nabla H\left( x \right)$  is the gradient of the function Hamilton. Equivalently,  Hamiltonian equations can be further written as:
$$\left\{ {{x}_{i}},H \right\}_{GPB}={{{J}_{ij}}\frac{\partial H}{\partial {{x}_{j}}}}$$

Let $M$ be a smooth manifold and let $s$ be a smooth real structural function on $M$ which is completely determined by the structure of manifold $M$.
Without loss of generality, we give a GSPB\footnote{GPB: Generalized Poisson bracket; GSPB: Generalized structural Poisson bracket; GCHS: Generalized Covariant Hamilton System; TGHS: thorough generalized Hamiltonian
system; PB: Poisson bracket; QCPB: Quantum Covariant Poisson Bracket;  GGC : generalized geometric commutator, } in a abstract covariant form,
\begin{definition}\label{d2} \cite{22}
  The generalized structural Poisson bracket of two functions $f,g\in {{C}^{\infty }}\left( M,\mathbb{R} \right)$ on $M$ is defined as
  $$\left\{ f,g \right\}={{\left\{ f,g \right\}}_{GPB}}+G\left(s, f,g \right)$$
  where $$G\left( s,f,g \right)=f{{\left\{ s ,g \right\}}_{GPB}}-g{{\left\{ s ,f \right\}}_{GPB}} =-G\left(s, g,f \right)$$ is called geometric bracket or geobracket, $s$ is  the structural function or the
geometric potential function.
\end{definition}
It is remarkable to see that the GSPB representation admits a dynamical geometric bracket formula instead of a generalized Poisson bracket structure.   we see that geobracket always satisfies the covariant condition $G\left(s, f,g \right)\neq 0$.
Figuratively,  the structural function or the
geometric potential function  $s$ is like a stage or the background for the movement of the objects.

Intuitively, one can understand this defined formulation by noting that the second part, geobracket, can be treated as the second part of the geodesic equation, and in so doing will make it clear. The GSPB depends smoothly on both ${{\left\{ f,g \right\}}_{GPB}}$ and the geobracket $G\left(s,f,g \right) $. Obviously, the geobracket $G\left(s,f,g \right) $ is necessary for a complete Hamiltonian syatem, as a result of the geobracket $G\left(s,f,g \right) $, it can be generally used to depict nonlinear system in a non-Euclidean space, and for general manifolds.   As GSPB defined, the GCHS is completely assured and determined by the GSPB. Therefore, the GCHS can be distinctly obtained.

\subsection{Generalized Covariant Hamilton System}
\begin{theorem}[GCHS] \cite{22}
The GCHS of two functions $f,H\in {{C}^{\infty }}\left( M,\mathbb{R} \right)$ on $M$ is defined as
\[ \frac{\mathcal{D}f}{dt}=\left\{ f,H \right\}={{\left\{ f,H \right\}}_{GPB}}+G\left(s,f,H \right) \]for ${{\mathbb{R}}^{r}}$,  the geobracket is
\begin{align}
 G\left(s,f,H \right) =f{{\left\{ s ,H \right\}}_{GPB}}-H{{\left\{ s ,f \right\}}_{GPB}} \notag
\end{align}in terms of $f$ and Hamiltonian $H$,  $s$ is structural function on $M$.
\end{theorem}In fact, general covariance of the GCHS holds naturally.
The GCHS is totally dependent on the GSPB and accordingly defined. Notice that the GCHS is covariant, it means that the GCHS is complete, and its form is invariant under the coordinate transformation, just like the covariance of the geodesic equation. So, the motion is completely determined by the structure of the manifold $M$.

More precisely, the generalized covariant Hamilton system includes two parts as follows:\\
The thorough generalized Hamiltonian system (TGHS)\newline~~~~~ $\frac{df}{dt}= {\dot {f}}={{\left\{ f,H \right\}}_{GPB}}-H{{\left\{ s ,f \right\}}_{GPB}}$,\newline
The S-dynamics:\newline $\frac{ds }{dt}=w={{\left\{ s ,H \right\}}_{GPB}}=\left\{ 1,H \right\}$,\newline
where $\frac{\mathcal{D}}{dt}=d/dt+w$ is the covariant time derivative.

By defining the GSPB which can basically solve and describe an entire Hamiltonian system. Thus, in order to state this point. To start with, we recall the basic notions of the GSPB. We explain that the GCHS corresponds to non-Euclidean space which corresponds to curved coordinate systems in curved space-time. The geobracket $G\left(s,f,H \right)$ is the part that has been neglected for a long time which is rightly the correction term for the general covariance which certainly means nonlinear system.  It is precisely because of the appearance of the second term  $G\left(s,f,H \right)$  that the GCHS can describe and explain the generalized nonlinear system, the nonlinear Hamiltonian system.

In the following, the covariant equilibrium equation of the GCHS as a special case is naturally generated as follows.
\begin{corollary}
 The covariant equilibrium equation of the GCHS is $\frac{\mathcal{D}f}{dt}=\left\{ f,H \right\}=0$, i.e, $${{\left\{ f,H \right\}}_{GPB}}+G\left(s,f,H \right)=0$$holds, then $f$ is called covariant
conserved quantity.
\end{corollary}
The critical points of the GCHS are precisely corresponding to the geodesics-like.
In an appropriate sense, zeros of the GCHS are thus regarded as function $f$ through geodesics-like.

For example, the formulations of the component of generalized covariant Hamilton system in terms of the coordinates are shown as follows,
$$\frac{\mathcal{D}{{x}_{k}}}{dt}=\left\{ {{x}_{k}},H\right\}={\dot{x}_{k}}+{{x}_{k}}w={{J}_{kj}}{{\partial}_{j}}H+H{{c}_{k}}+{{x}_{k}}w,$$
 $${\dot{x}_{k}}=\frac{d{{x}_{k}}}{dt}={{J}_{kj}}{{D}_{j}}H
 ={{J}_{kj}}{{\partial}_{j}}H+H{{c}_{k}},$$
\[w=\frac{ds}{dt}={{A}_{k}}{\dot{x}_{k}}={{b}_{j}}{{\partial }_{j}}H.\]where ${{c} _{j}}={{J}_{ji}}{{A}_{i}}=-b_{j}$, and ${A}_{j}={{\partial }_{j}}s$, note that the S-dynamics describes the rotations of the manifold in an angular frequency.

With the help of the structural function, the GSPB\footnote{Notes: GCC: Geometric canonical commutation; CCHE:canonical covariant Hamilton equations; GHE: generalized Heisenberg equations; CTHE: Canonical thorough Hamilton equations.}. is well defined for covariant generalized mechanics.
\begin{center}
\begin{tabular}{c r @{.} l}
\hline\hline
The covariant generalized mechanics  \\
\hline
GSPB:  $\left\{ f,g \right\}={{\left\{ f,~g \right\}}_{GPB}}+G\left(s,f,g \right)$, ~~  $G\left(s,f,g \right) =f{{\left\{ s ,g \right\}}_{GPB}}-g{{\left\{ s ,f \right\}}_{GPB}}$\\
\hline
CCHE: $\left\{ {{x}_{j}},~{{p}_{k}} \right\}={{\delta }_{jk}}+{{x}_{j}}{{\partial }_{k}}s+{{p}_{k}}{{J}_{jq}}{{\partial }_{q}}s$\\
\hline
Hamilton function:  $H\left( x,~p \right)$ \\
\hline
GCHS:   $\frac{\mathcal{D}{{x}_{k}}}{dt}=\left\{{{x}_{k}} ,H \right\}={\dot{x}_{k}}+{{x}_{k}}w,~~\frac{\mathcal{D}{{p}_{k}}}{dt}=\left\{{{p}_{k}} ,H\right\}=-{{D}_{k}}H+{{p}_{k}}w$
\\
\hline
CTHE:
${\dot{x}_{k}}={{J}_{kj}}{{D}_{j}}H,~~{\dot{p}_{k}}=-{{D}_{k}}H$
\\ \hline
S-dynamics: $\frac{ds }{dt}=w={{\left\{ s ,H \right\}}_{GPB}}=\left\{ 1,H \right\}$.
\\
\hline\hline
\end{tabular}
\end{center}
where $\left\{ f,g \right\}_{GPB}={{\nabla }^{T}}fJ\nabla g={{J}_{ij}}\frac{\partial f}{\partial {{x}_{i}}}\frac{\partial g}{\partial {{x}_{j}}}$ is generalized Poisson bracket.

Essentially, each different bracket corresponds to a Hamiltonian mechanical system, as a consequence,  HS with CPB, GHS and GPB, GCHS and GSPB. In \cite{22}, we have introduced an geometric approach to solve Hamiltonian mechanical problems based on the generalized structural Poisson bracket formulation. As a matter of fact, there always exists a structural function $s$ such that geometric bracket $G\left(s, f,g \right)$ holds for the GSPB and the GCHS on manifolds.

\section{Generalized geometric commutator and geomutator}
In this section, we shall deduce certain results that how generalized geometric commutator is defined generally.

Let $a$ and $b$ be any two operators, their commutator is formally defined by
 \cite{12,13,14,21}
${\displaystyle [a,b]_{cr}=ab-ba.}$
Note that the operator for commutator can be any mathematical form to be appeared for the calculation, such as a function, vector, differential operator, partial differential operator, even a number in a number field, and so on, it can be arbitrarily chosen according to our needs.

In linear algebra, if two endomorphisms of a space are represented by commuting matrices in terms of one basis, then they are so represented in terms of every basis. By using the commutator as a Lie bracket, every associative algebra can be turned into a Lie algebra.

The commutator has the following properties: \cite{12,13,14,21}
$${\displaystyle [a+b,c]_{cr}=[a,c]_{cr}+[b,c]_{cr}}$$
$${\displaystyle [a,b]_{cr}=-[b,a]_{cr}}$$
$${\displaystyle [a,[b,c]_{cr}]_{cr}+[b,[c,a]_{cr}]_{cr}+[c,[a,b]_{cr}]_{cr}=0}$$
The third relation is called anticommutativity, while the fourth is the Jacobi identity.

Furthermore, let's give some more specific examples for a better understanding above definition of the commutator.
\begin{enumerate}
\item The commutator is $\left( f,g \right)\mapsto {{\left[ \hat{f}, \hat{g} \right]}_{cr}}=\hat{f} \hat{g}-\hat{g}\hat{f}$ for the operators $\hat{f}, \hat{g}$.

  \item Let's consider the Lie bracket,
  $\left( X,Y \right)\mapsto {{\left[ X,Y \right]}_{cr}}=XY-YX\in  {{\mathfrak{X}}^{r}}\left( M \right)$,
where $X,Y\in {{\mathfrak{X}}^{r}}\left( M \right)$ and ${{\mathfrak{X}}^{r}}\left( M \right) $ means vector space\cite{19,20}.
\end{enumerate}

Let $a$ and $b$ be any elements of any algebra, or any operators, their generalized geometric commutator is formally defined by the following.
\begin{definition}\label{d4}
A generalized geometric commutator of arity two $a,b$ is formally given by
$$\left[ a,b \right]={{\left[ a,b \right]}_{cr}}+G\left( s,a,b \right)$$
The geomutator is $$G\left( s,a,b \right)=a{{\left[ s,b \right]}_{cr}}-b{{\left[ s,a \right]}_{cr}}$$satisfying $G\left( s,a,b \right)=-G\left( s,b,a\right)$, where $s$ is a geometric potential function or structural function given by domain.
\end{definition}
Note that a generalized geometric commutator (GGC) can be regarded as a geometric binary operation, a geometric binary operation may also involve several sets, that is, mixed set. In fact,  the generalized geometric commutator can be seen as an algorithm or operational rule for possible needs, this is a expression of highly formalism, namely, generalized geometric commutator is a deformation and extension of formalization relative to the commutator, we only consider this formalized structure for considering the application of it,  owing to this point, it can be used for further considerations. As a consequence, it can be used with no restrictions.
\begin{theorem}
  The generalized geometric commutator has the following properties:
  \begin{enumerate}
    \item $\left[ a,a \right]=0$
    \item $\left[ a,c \right]=-\left[ c,a \right]$
    \item $\left[ a+b,c \right]=\left[ a,c \right]+\left[ b,c \right]$
    \item $\left[ ab,c \right]=a{{\left[ b,c \right]}_{cr}}+{{\left[ a,c \right]}_{cr}}b+G\left( s,ab,c \right)$
    \item $\left[ a,\left[ b,c \right] \right]+\left[ b,\left[ c,a \right] \right]+\left[ c,\left[ a,b \right] \right]=0$
  \end{enumerate}
\end{theorem}
By modelling generalized geometric commutator, we can consider more mathematical structure and its applications for other fields.
More precisely, based on the formal expression, the geomutator above is accurately written in the form
\begin{align}
  G\left( s,a,b \right)&=asb-bsa+bas-abs=\left\langle a:s:b \right\rangle -\left[ a,b \right]_{cr}s \notag
\end{align}
where
$\left\langle a:s:b \right\rangle =asb-bsa $, and
 $ \left[ a,b \right]_{cr}s=abs-bas$.
Accordingly, the
generalized geometric commutator of arity two $a,b$ is formally rewritten by
$$\left[ a,b \right]={{\left[ a,b \right]}_{cr}}+\left\langle a:s:b \right\rangle -\left[ a,b \right]_{cr}s$$

\begin{definition}
 The covariant equilibrium equation is given by $\left[ a,b \right]=0$, i.e,
 ${{\left[ a,b \right]}_{cr}}+G\left( s, a,b \right)=0$ for arbitrary operators $a,~b$.
\end{definition}
Note that covariant equilibrium equation is also written as
$${{\left[ a,b \right]}_{cr}}+\left\langle a:s:b \right\rangle -\left[ a,b \right]_{cr}s=0$$ or ${{\left[ a,b \right]}_{cr}}=-G\left( s, a,b \right)$.
When generalized geometric commutator comes to the quantum operator, quantum covariant Poisson bracket appears for a complete extension of classical quantum Poisson bracket. In such case, the arity two $a,b$ will be replaced by specific quantum operator, in this applicable fields, the ${{\left[ a,b \right]}_{cr}}$ emerges in quantum mechanics as quantum Poisson bracket form ${{\left[ \hat{a},\hat{b} \right]}_{QPB}}$.

\subsection{Geomutator}
Some properties of the geomutator are given by
\begin{align}
  & G\left( s,a+b,c+d \right)=G\left( s,a,c \right)+G\left( s,b,d \right)+G\left( s,b,c \right)+G\left( s,a,d \right) \notag\\
 & G\left( s,a+b,c \right)=G\left( s,a,c \right)+G\left( s,b,c \right)\notag \\
 & G\left( s,a,c+d \right)=G\left( s,a,c \right)+G\left( s,a,d \right)\notag
\end{align}
for elements $a,b,c,d$. With some particular properties are given by
\begin{align}
  & G\left( s,a,a \right)=G\left( s,s,s \right)=0 \notag\\
 & G\left( s,s,a \right)=s{{\left[ s,a \right]}_{cr}} \notag\\
 & G\left( s,a,s \right)=s{{\left[ a,s \right]}_{cr}}\notag
\end{align}
As generalized geometric commutator stated above, by using the transformation \[a\to {{a}^{\left( s \right)}}=a+sa,~~b\to {{b}^{\left( s \right)}}=b+sb\]The generalized geometric commutator is transformed to
\begin{align}
 \left[ a,b \right] &={{\left[ a,b \right]}_{cr}}+G\left( s,a,b \right) \notag\\
 & =ab-ba+asb-bsa+bas-abs \notag\\
 & =a\left( b+sb \right)-b\left( a+sa \right)-{{\left[ a,b \right]}_{cr}}s \notag\\
 & =a{{b}^{\left( s \right)}}-b{{a}^{\left( s \right)}}-{{\left[ a,b \right]}_{cr}}s \notag
\end{align}
or taking the transformation
$a\to {{a}^{\left( sg \right)}}=a+sa-as$, and ${{\left[ a,s \right]}_{cr}}\neq 0$ holds generally,  then
\begin{align}
 \left[ a,b \right] & ={{\left[ a,b \right]}_{cr}}+G\left( s,a,b \right) =a\left( b-bs+sb \right)-b\left( a+sa-as \right) \notag\\
 & =a{{b}^{\left( sg \right)}}-b{{a}^{\left( sg \right)}} \notag
\end{align}
Obviously, we can easily check that generalized geometric commutator is as a generalization of commutator mainly reflected in the transformation of the elements.

\section{Quantum covariant Poisson bracket}
In this section, we start with the definition of quantum covariant Poisson bracket that generalizes the quantum Poisson bracket by using the structure function $s$ associated with the structure of space. We shall see that  quantum covariant Poisson bracket is defined by the generalized geometric commutator as definition \ref{d4} stated.  Once the framework of the quantum covariant Poisson bracket is established along with quantum geobracket \footnote{Quantum geometric bracket is abbreviated as quantum geobracket} defined by geomutator, then it can be continued to generally explain the quantum mechanics more precisely,  in the same way as was done for the GSPB obtained by the same procedure. The QCPB ${{\left[ \hat{f},\hat{g} \right]}_{QCPB}}\equiv \left[ \hat{f},\hat{g} \right]$ will be denoted throughout the paper.

\begin{definition}[QCPB]
  The QCPB is generally defined as
 \[\left[ \hat{f},\hat{g} \right]={{\left[ \hat{f},\hat{g} \right]}_{QPB}}+G\left( s,\hat{f},\hat{g} \right)\]in terms of quantum operator $\hat{f},~\hat{g}$, where $$G\left(s, \hat{f},\hat{g} \right)=\hat{f}{{\left[ s,\hat{g} \right]}_{QPB}}-\hat{g}{{\left[ s,\hat{f} \right]}_{QPB}}=-G\left(s, \hat{g},\hat{f} \right)$$ is called quantum geometric bracket.
\end{definition}
As defined above, $\left[ \cdot ,~\cdot  \right]={{\left[ \cdot ,~\cdot  \right]}_{QPB}}+G\left(s,\cdot ,~\cdot  \right)$ denotes the generalized geometric commutator (GGC) which is associated with  geometric potential function ( structural function) $s$ of two operators, note that geometric potential function $s$ can be treated as a special operator like the position operator $\hat{x}=x$, namely,  $\hat{s}=s$, and $G\left(\hat{s}, \hat{f},\hat{g} \right)=G\left(s, \hat{f},\hat{g} \right)$,  occasionally, the geometric potential function $s$ can be understood as the quantum system's own.  Therefore,    $G\left(s,~\cdot ,~\cdot  \right)$ is amendatory formula to the classical bracket, as a corrected term of the QCPB to the QPB.  It is zero if and only if $\hat{f}$ and $\hat{g}$ covariant commute, i,e. $\left[ \hat{f},\hat{g} \right]=0$.  It is remarkable to see that the QCPB representation admits a dynamical geometric bracket formula on the manifold. Note that  geometric potential function $s$ represents the background property of spacetime.

Note that the QCPB can be also treated as the generalized geometric commutator, somehow,  it has the following properties:
\begin{enumerate}
  \item $\left[ \hat{f}+\hat{h},\hat{g} \right]=\left[ \hat{f},\hat{g} \right]+\left[ \hat{h},\hat{g} \right]$.
  \item $\left[ \hat{f},\hat{g} \right]=-\left[ \hat{g},\hat{f} \right],\left[ \hat{f},\hat{f} \right]=0$.
  \item $\left[ \left[ \hat{f},\hat{g} \right],\hat{h} \right]+\left[ \left[ \hat{g},\hat{h} \right],\hat{f} \right]+\left[ \left[ \hat{h},\hat{f} \right],\hat{g} \right]=0$.
   \item
      $\left[ \hat{f},\hat{h}\hat{g} \right]
=\hat{h}\left[ \hat{f},\hat{g} \right]+\left[ \hat{f},\hat{h} \right]\hat{g}+{{\left[ \hat{f},\hat{h} \right]}_{QPB}}{{\left[ s,\hat{g} \right]}_{QPB}}+\hat{h}{{\left[ s,\hat{f} \right]}_{QPB}}\hat{g}$.
\end{enumerate}
Relation 2 is called anticommutativity, while 3 is the generalized Jacobi identity, identity 4 can be interpreted as a generalized Leibniz rule. Note that the generalized Jacobi identity has a complicated form
\begin{align}
 \left[ \left[ \hat{f},\hat{g} \right],\hat{h} \right] &=\left[ {{\left[ \hat{f},\hat{g} \right]}_{QPB}}+G\left( s,\hat{f},\hat{g} \right),\hat{h} \right]=\left[ {{\left[ \hat{f},\hat{g} \right]}_{QPB}},\hat{h} \right]+\left[ G\left( s,\hat{f},\hat{g} \right),\hat{h} \right] \notag\\
 & ={{\left[ \left[ \hat{f},\hat{g} \right],\hat{h} \right]}_{QPB}}+G\left( s,\left[ \hat{f},\hat{g} \right],\hat{h} \right) \notag\\
 & ={{\left[ {{\left[ \hat{f},\hat{g} \right]}_{QPB}}+G\left( s,\hat{f},\hat{g} \right), \hat{h} \right]}_{QPB}}+G\left( s,{{\left[ \hat{f},\hat{g} \right]}_{QPB}}+G\left( s,\hat{f},\hat{g} \right),\hat{h} \right) \notag
\end{align}
As a result, the generalized Jacobi identity is completely rewritten as
\begin{align}
 {{N}_{cl}}\left( \hat{f},\hat{g},\hat{h} \right) &=\left[ \left[ \hat{f},\hat{g} \right],\hat{h} \right]+\left[ \left[ \hat{g},\hat{h} \right],\hat{f} \right]+\left[ \left[ \hat{h},\hat{f} \right],\hat{g} \right] \notag\\
 & ={{N}_{cc}}\left( \hat{f},\hat{g},\hat{h} \right)+{{N}_{ll}}\left( s,\hat{f},\hat{g},\hat{h} \right)=0 \notag
\end{align}where the classical Jacobi identity is denoted as
\[{{N}_{cc}}\left( \hat{f},\hat{g},\hat{h} \right)={{\left[ {{\left[ \hat{f},\hat{g} \right]}_{QPB}},\hat{h} \right]}_{QPB}}+{{\left[ {{\left[ \hat{g},\hat{h} \right]}_{QPB}},\hat{f} \right]}_{QPB}}+{{\left[ {{\left[ \hat{h},\hat{f} \right]}_{QPB}},\hat{g} \right]}_{QPB}}\]
and
\begin{align}
  & {{N}_{ll}}\left( s,\hat{f},\hat{g},\hat{h} \right)=G\left( s,{{\left[ \hat{f},\hat{g} \right]}_{QPB}},\hat{h} \right)+{{\left[ G\left( s,\hat{f},\hat{g} \right),\hat{h} \right]}_{QPB}}+G\left( s,G\left( s,\hat{f},\hat{g} \right),\hat{h} \right) \notag\\
 & +G\left( s,{{\left[ \hat{g},\hat{h} \right]}_{QPB}},\hat{f} \right)+{{\left[ G\left( s,\hat{g},\hat{h} \right),\hat{f} \right]}_{QPB}}+G\left( s,G\left( s,\hat{g},\hat{h} \right),\hat{f} \right) \notag\\
 & +G\left( s,{{\left[ \hat{h},\hat{f} \right]}_{QPB}},\hat{g} \right)+{{\left[ G\left( s,\hat{h},\hat{f} \right),\hat{g} \right]}_{QPB}}+G\left( s,G\left( s,\hat{h},\hat{f} \right),\hat{g} \right) \notag
\end{align}
Notice that ${{\left[ \hat{f},\hat{g} \right]}_{QPB}}=\hat{f}\hat{g}-\hat{g}\hat{f}$, then quantum geobracket can be precisely expressed as
\begin{align}
  G\left(s,\hat{f},\hat{g} \right)&=\hat{f}{{\left[ s,\hat{g} \right]}_{QPB}}-\hat{g}{{\left[ s,\hat{f} \right]}_{QPB}} \notag\\
 & =\hat{f}\left( s\hat{g} \right)-\hat{g}\left( s\hat{f} \right)+\hat{g}\left( \hat{f}s \right)-\hat{f}\left( \hat{g}s \right) \notag\\
 & =-G\left(s,\hat{g},\hat{f} \right) \notag
\end{align}
that is a antisymmetric bracket dependent on the structural function $s$. Thus, the QCPB satisfies the antisymmetric property
\[\left[ \hat{f},\hat{g} \right]={{\left[ \hat{f},\hat{g} \right]}_{QPB}}+G\left(s,\hat{f},\hat{g} \right)=-{{\left[ \hat{g},\hat{f} \right]}_{QPB}}-G\left(s,\hat{g},\hat{f} \right)=-\left[ \hat{g},\hat{f} \right]\]
or the quantum geobracket is given by
\[G\left(s,\hat{f},\hat{g} \right)=\left\langle \hat{f}:s:\hat{g} \right\rangle -\left[ \hat{f},\hat{g} \right]s\]
where $\left\langle \hat{f}:s:\hat{g} \right\rangle =\hat{f}\left( s\hat{g} \right)-\hat{g}\left( s\hat{f} \right)$.

Above all, the QCPB is completely written as
\begin{align}
 \left[ \hat{f},\hat{g} \right] & ={{\left[ \hat{f},\hat{g} \right]}_{QPB}}+G\left( s,\hat{f},\hat{g} \right)\notag \\
 & ={{\left[ \hat{f},\hat{g} \right]}_{QPB}}+\hat{f}{{\left[ s,\hat{g} \right]}_{QPB}}-\hat{g}{{\left[ s,\hat{f} \right]}_{QPB}} \notag\\
  & ={{\left[ \hat{f},\hat{g} \right]}_{QPB}}+\left\langle \hat{f}:s:\hat{g} \right\rangle -\left[ \hat{f},\hat{g} \right]s\notag
\end{align}

\begin{definition}
 The covariant equilibrium equation is given by $\left[ \hat{f},\hat{g} \right]=0$, i.e,
 ${{\left[ \hat{f},\hat{g} \right]}_{QPB}}+G\left( s, \hat{f},\hat{g} \right)=0$ for operators $\hat{f},~\hat{g}$.
\end{definition}
The operator identity $$G\left(s, \hat{f},\hat{h}\hat{g} \right)=\hat{f}{{\left[ s,\hat{h}\hat{g} \right]}_{QPB}}-\hat{h}\hat{g}{{\left[ s,\hat{f} \right]}_{QPB}}$$
allows the evaluation of the commutator of $\hat{f},\hat{h}$ with $\hat{g}$,
\begin{align}
 G\left( s,\hat{f},\hat{h}\hat{g} \right) &=\hat{f}{{\left[ s,\hat{h}\hat{g} \right]}_{QPB}}-\hat{h}\hat{g}{{\left[ s,\hat{f} \right]}_{QPB}} \notag\\
 & =\hat{f}{{\left[ s,\hat{h} \right]}_{QPB}}\hat{g}+\hat{f}\hat{h}{{\left[ s,\hat{g} \right]}_{QPB}}-\hat{h}\hat{g}{{\left[ s,\hat{f} \right]}_{QPB}}\notag
\end{align}
where we have used ${{\left[ s,\hat{h}\hat{g} \right]}_{QPB}}={{\left[ s,\hat{h} \right]}_{QPB}}\hat{g}+\hat{h}{{\left[ s,\hat{g} \right]}_{QPB}}$,
 and it implies that
\[\left[ \hat{f},\hat{h}\hat{g} \right]=\hat{h}\left[ \hat{f},\hat{g} \right]+\left[ \hat{f},\hat{h} \right]\hat{g}+\eta \left( s,\hat{f},\hat{h},\hat{g} \right)\]
where
\[\eta \left( s,\hat{f},\hat{h},\hat{g} \right)={{\left[ \hat{f},\hat{h} \right]}_{QPB}}{{\left[ s,\hat{g} \right]}_{QPB}}+\hat{h}{{\left[ s,\hat{f} \right]}_{QPB}}\hat{g}\]

\subsection{Examples of QCPB}
We can give a example for better understanding of the QCPB, for instance,
if let $$\hat{f}=-\sqrt{-1}{{e}^{\sqrt{-1}x}}\frac{d}{dx},~~
\hat{g}={{e}^{\sqrt{-1}x}}$$ be given, then the classical operation leads to the consequence
\begin{align}
[\hat{f}, \hat{g}]_{QPB} \psi(x) &
=(\hat{f} \hat{g}-\hat{g} \hat{f}) \psi(x) \notag\\
&=e^{2\sqrt{-1}x} \psi(x)-\sqrt{-1}e^{2\sqrt{-1}x} \frac{d}{d x} \psi(x)+\sqrt{-1} e^{2\sqrt{-1}x} \frac{d}{d x} \psi(x) \notag\\
&=e^{2\sqrt{-1} x} \psi(x)\notag
\end{align}
Then $[\hat{f}, \hat{g}]_{QPB}=e^{2\sqrt{-1} x}$.
But by using the QCPB, we only need to evaluate quantum geometric bracket $G\left( s,\hat{f},\hat{g} \right)$. More precisely,
\begin{align}
 \hat{f}{{\left[ s,\hat{g} \right]}_{QPB}}\psi &=-\sqrt{-1}{{e}^{\sqrt{-1}x}}\frac{d}{dx}\left( s{{e}^{\sqrt{-1}x}}\psi -{{e}^{\sqrt{-1}x}}s\psi  \right)=0\notag \\
 & \hat{g}{{\left[ s,\hat{f} \right]}_{QPB}}\psi =\psi \sqrt{-1}{{e}^{2\sqrt{-1}x}}\frac{ds}{dx}=-\psi {{e}^{\sqrt{-1}x}}\hat{f}s \notag
\end{align}As a result of above computation, we get
$\hat{g}{{\left[ s,\hat{f} \right]}_{QPB}}=-{{e}^{\sqrt{-1}x}}\hat{f}s$.
And the quantum geometric bracket is finally shown as
\begin{align}
 G\left( s,\hat{f},\hat{g} \right)\psi &=\hat{f}{{\left[ s,\hat{g} \right]}_{QPB}}\psi -\hat{g}{{\left[ s,\hat{f} \right]}_{QPB}}\psi  \notag\\
 & =-\hat{g}{{\left[ s,\hat{f} \right]}_{QPB}}\psi =\psi {{e}^{\sqrt{-1}x}}\hat{f}s \notag\\
 & =-\psi \sqrt{-1}{{e}^{2\sqrt{-1}x}}\frac{ds}{dx} \notag
\end{align}
Accordingly, the QCPB is obtained
\begin{align}
 \left[ \hat{f},\hat{g} \right] &={{e}^{2\sqrt{-1}x}}\left( 1-\sqrt{-1}\frac{ds}{dx} \right) \notag
\end{align}
We can check that if $\frac{ds}{dx} =0$ is given, then
$\left[ \hat{f},\hat{g} \right]$ is degenerated to the classical case $ {{\left[ \hat{f},\hat{g} \right]}_{QPB}}$, equivalently, $G\left( s,\hat{f},\hat{g} \right)=0$ disappears.

For another example,  let $\hat{f}=\frac{d}{dx},
\hat{g}=x$ be given, and wave function is $\psi \left( x \right)=\sin x$, then
the QCPB is given by
\begin{align}
 \left[ \frac{d}{dx},x \right]\psi \left( x \right) &={{\left[ \frac{d}{dx},x \right]}_{QPB}}\sin x+G\left( s,\frac{d}{dx},x \right)\sin x  \notag
\end{align}
The QPB and the quantum geometric bracket can be respectively evaluated by
${{\left[ \frac{d}{dx},x \right]}_{QPB}}\sin x=\sin x$,
and
 $G\left( s,\frac{d}{dx},x \right)\sin x =x\sin x\frac{ds}{dx}$, then it leads to the outcome
\[\left[ \frac{d}{dx},x \right]\psi \left( x \right)=\left( 1+x\frac{ds}{dx} \right)\psi \left( x \right)\]

\subsection{Quantum geometric bracket}
As previously stated, it turns out that the quantum geometric bracket is given by
\[G\left( s,\hat{f},\hat{g} \right)=\left[ \hat{f},\hat{g} \right]-{{\left[ \hat{f},\hat{g} \right]}_{QPB}}\]Actually, it's a map expressed as
$\hat{f},\hat{g}\to G\left(s, \hat{f},\hat{g} \right)$ with respect to two quantum operators $\hat{f},~\hat{g}$.
In general, in order to preserve the antisymmetry of the QPB, the quantum geometric bracket is always asked to satisfy the antisymmetry, i,e.
$G\left(s, \hat{f},\hat{g} \right)=- G\left( s,\hat{g},\hat{f} \right)$.
Thus, let's consider the properties of quantum geobracket, more precisely,  $$G\left(s,\hat{f},\hat{g} \right)=\hat{f}{{\left[ s,\hat{g} \right]}_{QPB}}-\hat{g}{{\left[ s,\hat{f} \right]}_{QPB}}$$ that fulfils the antisymmetry $G\left(s,\hat{f},\hat{g} \right)=-G\left(s, \hat{g},\hat{f} \right)$, then some special properties of the quantum geometric bracket can be represented by the following
\begin{align}
  & G\left(s,\hat{f},\hat{f} \right)=G\left(s,s,s \right)=0  \notag\\
 & G\left(s,s,\hat{g} \right)=s{{\left[ s,\hat{g} \right]}_{QPB}}  \notag\\
 & G\left(s,\hat{f},s \right)=s{{\left[ \hat{f},s \right]}_{QPB}} \notag
\end{align}
The QCPB has greatly enlarged the research scope of canonical commutation relation, obviously, it completes and improves the quantum Poisson bracket. In short, the QCPB is completely written as
\begin{align}
 \left[ \hat{f},\hat{g} \right] &={{\left[ \hat{f},\hat{g} \right]}_{QPB}}+\hat{f}{{\left[ s,\hat{g} \right]}_{QPB}}-\hat{g}{{\left[ s,\hat{f} \right]}_{QPB}} \notag\\
 & =\hat{f}\hat{g}-\hat{g}\hat{f}+\hat{f}\left( s\hat{g} \right)-\hat{g}\left( s\hat{f} \right)+\hat{g}\left( \hat{f}s \right)-\hat{f}\left( \hat{g}s \right) \notag
\end{align}
Thus, for a given wave function $\varphi$ onto the QCPB, it gives a generalized commutation relation
\begin{align}
 \left[ \hat{f},\hat{g} \right]\varphi &= {{\left[ \hat{f},\hat{g} \right]}_{QPB}}\varphi +\hat{f}{{\left[ s,\hat{g} \right]}_{QPB}}\varphi -\hat{g}{{\left[ s,\hat{f} \right]}_{QPB}}\varphi  \notag\\
 & =\hat{f}\hat{g}\varphi -\hat{g}\hat{f}\varphi +\hat{f}\left( s\hat{g}\varphi  \right)-\hat{g}\left( s\hat{f}\varphi  \right)+\hat{g}\hat{f}\left( s\varphi  \right)-\hat{f}\hat{g}\left( s\varphi  \right) \notag\\
 & ={{\left[ \hat{f},\hat{g} \right]}_{QPB}}\varphi +{{\left[ \hat{g},\hat{f} \right]}_{QPB}}\left( s\varphi  \right)+\hat{f}\left( s\hat{g}\varphi  \right)-\hat{g}\left( s\hat{f}\varphi  \right) \notag
\end{align}
As one can see, the QCPB has largely enriched the QPB for a better understanding of quantum mechanics, it clearly appears more parts such as
\[G\left(s,\hat{f},\hat{g} \right)=\hat{f}{{\left[ s,\hat{g} \right]}_{QPB}}-\hat{g}{{\left[ s,\hat{f} \right]}_{QPB}}=\hat{f}\left( s\hat{g} \right)-\hat{g}\left( s\hat{f} \right)+\hat{g}\left( \hat{f}s \right)-\hat{f}\left( \hat{g}s \right)\]This terms never emerges in the quantum mechanics for general thinking, therefore, the quantum geobracket $G\left( s, \hat{f},\hat{g} \right)$ is rightly a correction to remedy the QPB.

To consider the corresponding equilibrium of the QCPB, namley, $\left[ \hat{f},\hat{g} \right]=0$, in other words,
${{\left[ \hat{f},\hat{g} \right]}_{QPB}}+G\left( s, \hat{f},\hat{g} \right)=0$.
The QCPB has following properties
\begin{align}
  & \left[ \hat{f},\hat{g} \right]=-\left[ \hat{g},\hat{f} \right],~~\left[ \hat{f},\hat{f} \right]=0  \notag\\
 & \left[ \hat{f},\hat{h}+\hat{g} \right]=\left[ \hat{f},\hat{g} \right]+\left[ \hat{f},\hat{h} \right]  \notag\\
 & \left[ \hat{h}+\hat{g},\hat{f} \right]=\left[ \hat{g},\hat{f} \right]+\left[ \hat{h},\hat{f} \right]  \notag\\
 & \left[ \hat{f},c\hat{g} \right]=c\left[ \hat{f},\hat{g} \right] \notag \\
 & \left[ \hat{f},\hat{h}\hat{g} \right]=\hat{h}\left[ \hat{f},\hat{g} \right]+\left[ \hat{f},\hat{h} \right]\hat{g}+\eta \left( s,\hat{f},\hat{h},\hat{g} \right) \notag
\end{align}for any quantum operators $\hat{f},\hat{h},\hat{g}$,
where $c$ is a complex constant, and
\[\eta \left( s,\hat{f},\hat{h},\hat{g} \right)={{\left[ \hat{f},\hat{h} \right]}_{QPB}}{{\left[ s,\hat{g} \right]}_{QPB}}+\hat{h}{{\left[ s,\hat{f} \right]}_{QPB}}\hat{g}\]

Let
$\hat{f}={{\hat{f}}_{+}}+\sqrt{-1}{{\hat{f}}_{-}},~~
\hat{g}={{\hat{g}}_{+}}+\sqrt{-1}{{\hat{g}}_{-}}$
be given for generally calculations of the QCPB, where ${{\hat{f}}_{+}},{{\hat{f}}_{-}},{{\hat{g}}_{+}},{{\hat{g}}_{-}}$ are Hermitian operators.
\begin{align}
 \left[ \hat{f},\hat{g} \right] &=\left[ {{{\hat{f}}}_{+}}+\sqrt{-1}{{{\hat{f}}}_{-}},{{{\hat{g}}}_{+}}+\sqrt{-1}{{{\hat{g}}}_{-}} \right] \notag\\
 & ={{\left[ {{{\hat{f}}}_{+}}+\sqrt{-1}{{{\hat{f}}}_{-}},{{{\hat{g}}}_{+}}+\sqrt{-1}{{{\hat{g}}}_{-}} \right]}_{QPB}}+G\left( s,{{{\hat{f}}}_{+}}+\sqrt{-1}{{{\hat{f}}}_{-}},{{{\hat{g}}}_{+}}+\sqrt{-1}{{{\hat{g}}}_{-}} \right) \notag
\end{align}As a consequence, we only need to compute two parts of the QCPB respectively, the QPB and the quantum geometric bracket. To begin with the QPB is given by
\begin{align}
  & {{\left[ {{{\hat{f}}}_{+}}+\sqrt{-1}{{{\hat{f}}}_{-}},{{{\hat{g}}}_{+}}+\sqrt{-1}{{{\hat{g}}}_{-}} \right]}_{QPB}}={{\left[ {{{\hat{f}}}_{+}},{{{\hat{g}}}_{+}} \right]}_{QPB}}+{{\left[ {{{\hat{f}}}_{+}},\sqrt{-1}{{{\hat{g}}}_{-}} \right]}_{QPB}}\notag \\
 & \begin{matrix}
   {} & {} & {}  \\
\end{matrix}+{{\left[ \sqrt{-1}{{{\hat{f}}}_{-}},{{{\hat{g}}}_{+}} \right]}_{QPB}}+{{\left[ \sqrt{-1}{{{\hat{f}}}_{-}},\sqrt{-1}{{{\hat{g}}}_{-}} \right]}_{QPB}} \notag\\
 & ={{\left[ {{{\hat{f}}}_{+}},{{{\hat{g}}}_{+}} \right]}_{QPB}}-{{\left[ {{{\hat{f}}}_{-}},{{{\hat{g}}}_{-}} \right]}_{QPB}}+\sqrt{-1}\left( {{\left[ {{{\hat{f}}}_{-}},{{{\hat{g}}}_{+}} \right]}_{QPB}}+{{\left[ {{{\hat{f}}}_{+}},{{{\hat{g}}}_{-}} \right]}_{QPB}} \right)\notag
\end{align}
Subsequently, the quantum geometric bracket follows
\begin{align}
  & G\left( s,{{{\hat{f}}}_{+}}+\sqrt{-1}{{{\hat{f}}}_{-}},{{{\hat{g}}}_{+}}+\sqrt{-1}{{{\hat{g}}}_{-}} \right) \notag \\
 & =\left( {{{\hat{f}}}_{+}}+\sqrt{-1}{{{\hat{f}}}_{-}} \right){{\left[ s,{{{\hat{g}}}_{+}}+\sqrt{-1}{{{\hat{g}}}_{-}} \right]}_{QPB}}-\left( {{{\hat{g}}}_{+}}+\sqrt{-1}{{{\hat{g}}}_{-}} \right){{\left[ s,{{{\hat{f}}}_{+}}+\sqrt{-1}{{{\hat{f}}}_{-}} \right]}_{QPB}} \notag \\
 & ={{{\hat{f}}}_{+}}{{\left[ s,{{{\hat{g}}}_{+}}+\sqrt{-1}{{{\hat{g}}}_{-}} \right]}_{QPB}}-{{{\hat{g}}}_{+}}{{\left[ s,{{{\hat{f}}}_{+}}+\sqrt{-1}{{{\hat{f}}}_{-}} \right]}_{QPB}} \notag \\
 & \begin{matrix}
   {} & {} & {} & {}  \\
\end{matrix}+\sqrt{-1}\left\{ {{{\hat{f}}}_{-}}{{\left[ s,{{{\hat{g}}}_{+}}+\sqrt{-1}{{{\hat{g}}}_{-}} \right]}_{QPB}}-{{{\hat{g}}}_{-}}{{\left[ s,{{{\hat{f}}}_{+}}+\sqrt{-1}{{{\hat{f}}}_{-}} \right]}_{QPB}} \right\}\notag  \\
 & ={{{\hat{f}}}_{+}}{{\left[ s,{{{\hat{g}}}_{+}} \right]}_{QPB}}-{{{\hat{f}}}_{-}}{{\left[ s,{{{\hat{g}}}_{-}} \right]}_{QPB}}-{{{\hat{g}}}_{+}}{{\left[ s,{{{\hat{f}}}_{+}} \right]}_{QPB}}+{{{\hat{g}}}_{-}}{{\left[ s,{{{\hat{f}}}_{-}} \right]}_{QPB}} \notag \\
 & +\sqrt{-1}\left\{ {{{\hat{f}}}_{+}}{{\left[ s,{{{\hat{g}}}_{-}} \right]}_{QPB}}+{{{\hat{f}}}_{-}}{{\left[ s,{{{\hat{g}}}_{+}} \right]}_{QPB}}-{{{\hat{g}}}_{+}}{{\left[ s,{{{\hat{f}}}_{-}} \right]}_{QPB}}-{{{\hat{g}}}_{-}}{{\left[ s,{{{\hat{f}}}_{+}} \right]}_{QPB}} \right\} \notag
\end{align}Hence, we obtain the QCPB
\begin{align}
  & \left[ \hat{f},\hat{g} \right]=\left[ {{{\hat{f}}}_{+}}+\sqrt{-1}{{{\hat{f}}}_{-}},{{{\hat{g}}}_{+}}+\sqrt{-1}{{{\hat{g}}}_{-}} \right] \notag\\
 & ={{\left[ {{{\hat{f}}}_{+}},{{{\hat{g}}}_{+}} \right]}_{QPB}}+{{{\hat{f}}}_{+}}{{\left[ s,{{{\hat{g}}}_{+}} \right]}_{QPB}}-{{{\hat{g}}}_{+}}{{\left[ s,{{{\hat{f}}}_{+}} \right]}_{QPB}} \notag\\
 & \begin{matrix}
   {} & {} & {} & {}  \\
\end{matrix}-{{\left[ {{{\hat{f}}}_{-}},{{{\hat{g}}}_{-}} \right]}_{QPB}}-{{{\hat{f}}}_{-}}{{\left[ s,{{{\hat{g}}}_{-}} \right]}_{QPB}}+{{{\hat{g}}}_{-}}{{\left[ s,{{{\hat{f}}}_{-}} \right]}_{QPB}} \notag\\
 & +\sqrt{-1}\left\{ {{\left[ {{{\hat{f}}}_{-}},{{{\hat{g}}}_{+}} \right]}_{QPB}}+{{{\hat{f}}}_{-}}{{\left[ s,{{{\hat{g}}}_{+}} \right]}_{QPB}}-{{{\hat{g}}}_{+}}{{\left[ s,{{{\hat{f}}}_{-}} \right]}_{QPB}} \right. \notag\\
 & \begin{matrix}
   {} & {} & {} & {}  \\
\end{matrix}\left. +{{\left[ {{{\hat{f}}}_{+}},{{{\hat{g}}}_{-}} \right]}_{QPB}}+{{{\hat{f}}}_{+}}{{\left[ s,{{{\hat{g}}}_{-}} \right]}_{QPB}}-{{{\hat{g}}}_{-}}{{\left[ s,{{{\hat{f}}}_{+}} \right]}_{QPB}} \right\} \notag\\
 & =\left[ {{{\hat{f}}}_{+}},{{{\hat{g}}}_{+}} \right]-\left[ {{{\hat{f}}}_{-}},{{{\hat{g}}}_{-}} \right]+\sqrt{-1}\left( \left[ {{{\hat{f}}}_{-}},{{{\hat{g}}}_{+}} \right]+\left[ {{{\hat{f}}}_{+}},{{{\hat{g}}}_{-}} \right] \right) \notag
\end{align}
where
\begin{align}
 \left[ {{{\hat{f}}}_{+}},{{{\hat{g}}}_{+}} \right] &={{\left[ {{{\hat{f}}}_{+}},{{{\hat{g}}}_{+}} \right]}_{QPB}}+G\left( s,{{{\hat{f}}}_{+}},{{{\hat{g}}}_{+}} \right) \notag\\
 & ={{\left[ {{{\hat{f}}}_{+}},{{{\hat{g}}}_{+}} \right]}_{QPB}}+{{{\hat{f}}}_{+}}{{\left[ s,{{{\hat{g}}}_{+}} \right]}_{QPB}}-{{{\hat{g}}}_{+}}{{\left[ s,{{{\hat{f}}}_{+}} \right]}_{QPB}} \notag
\end{align}
Similarly,
\begin{align}
 \left[ \hat{f},\hat{g} \right] &=\left[ {{{\hat{f}}}_{+}}+\sqrt{-1}{{{\hat{f}}}_{-}},{{{\hat{g}}}_{+}}+\sqrt{-1}{{{\hat{g}}}_{-}} \right] \notag\\
 & ={{\left[ {{{\hat{f}}}_{+}},{{{\hat{g}}}_{+}} \right]}_{QPB}}-{{\left[ {{{\hat{f}}}_{-}},{{{\hat{g}}}_{-}} \right]}_{QPB}}+\sqrt{-1}\left( {{\left[ {{{\hat{f}}}_{-}},{{{\hat{g}}}_{+}} \right]}_{QPB}}+{{\left[ {{{\hat{f}}}_{+}},{{{\hat{g}}}_{-}} \right]}_{QPB}} \right) \notag\\
 & +G\left( s,{{{\hat{f}}}_{+}},{{{\hat{g}}}_{+}} \right)-G\left( s,{{{\hat{f}}}_{-}},{{{\hat{g}}}_{-}} \right)+\sqrt{-1}\left( G\left( s,{{{\hat{f}}}_{-}},{{{\hat{g}}}_{+}} \right)+G\left( s,{{{\hat{f}}}_{+}},{{{\hat{g}}}_{-}} \right) \right) \notag\\
 & ={{\left[ {{{\hat{f}}}_{+}}+\sqrt{-1}{{{\hat{f}}}_{-}},{{{\hat{g}}}_{+}}+\sqrt{-1}{{{\hat{g}}}_{-}} \right]}_{QPB}}+G\left( s,{{{\hat{f}}}_{+}}+\sqrt{-1}{{{\hat{f}}}_{-}},{{{\hat{g}}}_{+}}+\sqrt{-1}{{{\hat{g}}}_{-}} \right) \notag
\end{align}It implies that
\begin{align}
  & {{\left[ {{{\hat{f}}}_{+}}+\sqrt{-1}{{{\hat{f}}}_{-}},{{{\hat{g}}}_{+}}+\sqrt{-1}{{{\hat{g}}}_{-}} \right]}_{QPB}} \notag\\
 & ={{\left[ {{{\hat{f}}}_{+}},{{{\hat{g}}}_{+}} \right]}_{QPB}}-{{\left[ {{{\hat{f}}}_{-}},{{{\hat{g}}}_{-}} \right]}_{QPB}}+\sqrt{-1}\left( {{\left[ {{{\hat{f}}}_{-}},{{{\hat{g}}}_{+}} \right]}_{QPB}}+{{\left[ {{{\hat{f}}}_{+}},{{{\hat{g}}}_{-}} \right]}_{QPB}} \right) \notag\\
 & G\left( s,{{{\hat{f}}}_{+}}+\sqrt{-1}{{{\hat{f}}}_{-}},{{{\hat{g}}}_{+}}+\sqrt{-1}{{{\hat{g}}}_{-}} \right) \notag\\
 & =G\left( s,{{{\hat{f}}}_{+}},{{{\hat{g}}}_{+}} \right)-G\left( s,{{{\hat{f}}}_{-}},{{{\hat{g}}}_{-}} \right)+\sqrt{-1}\left( G\left( s,{{{\hat{f}}}_{-}},{{{\hat{g}}}_{+}} \right)+G\left( s,{{{\hat{f}}}_{+}},{{{\hat{g}}}_{-}} \right) \right)\notag
\end{align}
Above derivation reveals that the quantum geobracket as a new quantum geometric bracket remains the same operation as the QPB does. In other words, the operation rule of the quantum geobracket is parallel to the operation rule of the QPB.

\section{Quantum covariant Hamiltonian system}

This section will give the covariant dynamics which contains two different dynamics: the generalized Heisenberg equation and G-dynamics. It tells that the generalized Heisenberg equation perfects the Heisenberg equation by considering the structural function $s$. Firstly, let's give the definition of quantum covariant Hamiltonian system (QCHS) defined by the QCPB.

In the beginning, let us start with the QCPB by taking $\hat{g}=\hat{H}$ into consideration, then it formally yields below definition.
\begin{definition}[QCHS]
  The QCHS defined by QCPB in terms of quantum operator $\hat{f},~\hat{H}$ is generally given by
 \[\left[ \hat{f},\hat{H} \right]={{\left[ \hat{f},\hat{H} \right]}_{QPB}}+G\left(s, \hat{f},\hat{H} \right)\]where $$G\left(s, \hat{f},\hat{H} \right)=\hat{f}{{\left[ s,\hat{H} \right]}_{QPB}}-\hat{H}{{\left[ s,\hat{f} \right]}_{QPB}}$$ is quantum geobracket .
\end{definition}
Manifestly,
this operator $\hat{f}$ covariantly commutes with $H$.  By our construction, the QCHS derived from the QCPB is a transition for the further development of a complete and covariant theory that can naturally generalize the Heisenberg equation. Since $\hat{f}$ is an observable, $s$ is assumed to be real function as well. Our goal here is to give
a description of the method in general for perfecting the known theory such as Heisenberg equation. With that in mind, we proceed accordingly.

\begin{definition}\label{d5}
 The covariant equilibrium equation is given by $\left[ \hat{f},\hat{H} \right]=0$, i.e,
  ${{\left[ \hat{f},\hat{H} \right]}_{QPB}}+G\left( s, \hat{f},\hat{H} \right)=0$ for operators $\hat{f},~\hat{H}$, then $\hat{f}$ is called quantum covariant conserved quantity.
\end{definition}
Obviously, there is a certain case given by
\begin{equation} \label{eq4}
  \left[ \hat{H},\hat{H} \right]=-\hat{H}{{\left[ s,\hat{H} \right]}_{QPB}}+\hat{H}{{\left[ s,\hat{H} \right]}_{QPB}}=0
\end{equation}when $\hat{f}=\hat{H}$ based on definition \ref{d5}. It is clear to see that $\hat{H}$ is a quantum covariant conserved quantity.

\subsection{Covariant dynamics, generalized Heisenberg equation, G-dynamics}
We now illustrate this definition with one theorem below. More precisely,
the time covariant evolution of any observable $\hat{f}$ in the covariant dynamics is given by both generalized Heisenberg equation of motion and G-dynamics.

Based on the quantization map $Q$ mapping a function $f$ on the classical phase space to an operator ${\displaystyle Q_{f}}$ on the quantum Hilbert space, namely, $f\mapsto {{Q}_{f}}\equiv \hat{f}$.
Similarly, it goes for the GSPB and QCPB such that
$${\displaystyle Q_{\{f,g\}}={\frac {1}{\sqrt{-1}\hbar }}[Q_{f},Q_{g}]}$$
or conversely, $$[Q_{f},Q_{g}]=\sqrt{-1}\hbar{\displaystyle Q_{\{f,g\}}} $$
According to the rule of supplanting Poisson brackets by commutators written by Dirac in his book as Eq\eqref{eq2} shown. Similarly, we have same generally corresponding relation
\begin{align}
  & \left\{ f,g \right\}={{\left\{ f,g \right\}}_{GPB}}+G\left(s,f,g \right)  \notag\\
 & \mapsto \frac{1}{\sqrt{-1}\hbar }\left[ \hat{f},\hat{g} \right]=\frac{1}{\sqrt{-1}\hbar }{{\left[ \hat{f},\hat{g} \right]}_{QPB}}+\frac{1}{\sqrt{-1}\hbar }G\left(s,\hat{f},\hat{g} \right) \notag
\end{align}Here the geometric potential
function $s$ behaves like the position operator, in other words,  $\hat{s}=s$.
As a generalization to the manifolds, it certainly fits the curved spacetime which means that $\frac{1}{\sqrt{-1}\hbar }\left[ \hat{f},\hat{g} \right]$ is covariant, and quantum term $G\left(s,\hat{f},\hat{g} \right)$ is necessary to exist for a complete quantum theory.

As \eqref{eq8} did, analogously, this method of work lead us to propose that the quantum counterparts ${\hat {f}},{\hat {g}}$ of classical observables $f, g$ satisfy covariant rule \[\left[\hat{f},\hat{g}\right]=\sqrt{-1}\hbar \widehat{\left\{ f, g\right\}}\]where $\left[\hat{f},\hat{g}\right]$ is the QCPB, while $\left\{ f, g\right\}$ is the GSPB. More exactly,
\[{{\left[ \hat{f},\hat{g} \right]}_{QPB}}+G\left(s, \hat{f},\hat{g} \right)= \sqrt{-1}\hbar \widehat{\left( {{\left\{ f,g \right\}}_{GPB}}+G\left(s, f,g \right) \right)}\]
Actually, in order for the model to work for the quantum mechanics and elementary particles, we need to address specifically as follows:

\begin{theorem}\label{t2}
The covariant dynamics, the generalized Heisenberg equation, G-dynamics can be formally formulated as
\begin{description}
  \item[covariant dynamics:] $\frac{\mathcal{D}\hat{f}}{dt}=\frac{1}{\sqrt{-1}\hbar }\left[ \hat{f},\hat{H} \right]$
  \item[generalized Heisenberg equation:]
   $$\frac{d\hat{f}}{dt}=\frac{1}{\sqrt{-1}\hbar }{{\left[ \hat{f},\hat{H} \right]}_{QPB}}-\frac{1}{\sqrt{-1}\hbar }\hat{H}{{\left[ s,\hat{f} \right]}_{QPB}}$$
  \item[G-dynamics:] $\hat{w}=\frac{1}{\sqrt{-1}\hbar }{{\left[ s,\hat{H} \right]}_{QPB}}$.
\end{description}respectively, where $\frac{\mathcal{D}}{dt}=\frac{d}{dt}+\hat{w}$ is covariant time operator, and
${{\hat{H}}}$ is the Hamiltonian and $[\cdot,\cdot]$ denotes the GGC of two operators.
\end{theorem}Here, G-dynamics is only for the description of some kind of rotation, it's truly a rotation operator which never appears before in quantum mechanics, the G-dynamics is actually a frequency operator for describing the frequency of the manifold space or the environment.
Actually, quantum dynamical variables $\hat{f}$ covariantly evolves in time
\begin{align}
 \frac{\mathcal{D}}{dt}\hat{f} &=\frac{1}{\sqrt{-1}\hbar }\left[ \hat{f},\hat{H} \right] =\frac{1}{\sqrt{-1}\hbar }{{\left[ \hat{f},\hat{H} \right]}_{QPB}}-\frac{1}{\sqrt{-1}\hbar }\hat{H}{{\left[ s,\hat{f} \right]}_{QPB}} \notag\\
 & \begin{matrix}
   {} & {} & {}& {} & {}& {} & {}& {} & {}   \\
\end{matrix}+\frac{1}{\sqrt{-1}\hbar }\hat{f}{{\left[ s,\hat{H} \right]}_{QPB}} \notag\\
 & =\frac{d}{dt}\hat{f}+\hat{f}\hat{w} \notag
\end{align}
Analogously, this is similar to the dynamic equation in ordinary differential equation when covariant equilibrium $\frac{\mathcal{D}}{dt}\hat{f}=0$ holds, it also reveals that covariant equation $\frac{d}{dt}\hat{f}+\hat{f}\hat{w}=0$ is a complete expression in quantum mechanics. Thus, any observable that covariant commutes with the Hamiltonian usually is not a constant of the motion. It can be seen that the
G-dynamics is the quantum correspondence of the S-dynamics, that is to say,

 S-dynamics: $w={{\left\{ s ,H \right\}}_{GPB}}\rightarrow$ G-dynamics:
 $\hat{w}=\frac{1}{\sqrt{-1}\hbar }{{\left[ s,\hat{H} \right]}_{QPB}}$.

\subsection{Imaginary geomenergy}
In this section, the imaginary geomenergy \footnote{geometric energy is abbreviated as geomenergy} is defined based on the G-dynamics,  by using these new concepts, we can better give a presentation for the covariant dynamics, ect.
\begin{definition}\label{d6}
Based on the G-dynamics $\hat{w}$, then the imaginary geomenergy is induced by ${{E}^{\left( \operatorname{Im} \right)}}\left( \hat{w} \right)=\sqrt{-1}\hbar \hat{w}={{\left[ s,\hat{H} \right]}_{QPB}}$.
\end{definition}
Thus, a dynamical variable $\hat{f}$, which corresponds to a fixed linear operator, corresponds to a moving linear operator in this new scheme. It is clear that the covariant dynamics leads us to a complete quantum system in which the state of the system is represented by two separated dynamic quantum system, and the dynamical variables are represented by corresponding moving operators. This covariant dynamics which is outlined in another form below
\begin{align}
 \sqrt{-1}\hbar \frac{\mathcal{D}}{dt}\hat{f} &=\left[ \hat{f},\hat{H} \right] ={{\left[ \hat{f},\hat{H} \right]}_{QPB}}+G\left( s, \hat{f},\hat{H} \right) \notag\\
 & =\sqrt{-1}\hbar \frac{d}{dt}\hat{f}+\sqrt{-1}\hbar \hat{f}\hat{w} \notag
\end{align}
Based on the formula of G-dynamics, the covariant dynamics can be rewritten as
$$\frac{\mathcal{D}}{dt}\hat{f}=\frac{1}{\sqrt{-1}\hbar }\left[ \hat{f},\hat{H} \right]=\frac{d}{dt}\hat{f}+\hat{f}\hat{w}$$
where $\hat{w}$ is G-dynamics.
We consider the covariant equilibrium equation, that is, $\frac{\mathcal{D}}{dt}\hat{f}=0$, in other words, it's an equation shown by
$\frac{d}{dt}\hat{f}+\hat{f}\hat{w}=0$.
With the imaginary geomenergy defined above \ref{d6}, then covariant dynamics is rewritten in the form
\[\sqrt{-1}\hbar \frac{\mathcal{D}}{dt}\hat{f}=\left[ \hat{f},\hat{H} \right]=\sqrt{-1}\hbar \frac{d}{dt}\hat{f}+\hat{f}{{E}^{\left( \operatorname{Im} \right)}}\left( \hat{w} \right)\]

According to equation \eqref{eq4}, the covariant dynamics of motion in terms of Hamilton $\hat{H}$ can be written as
\[\frac{\mathcal{D}}{dt}\hat{H}=\frac{1}{\sqrt{-1}\hbar }\left[ \hat{H},\hat{H} \right]=0\]
Accordingly, the generalized Heisenberg equation follows
$\frac{d}{dt}\hat{H}=-\hat{H}\hat{w}$.

Obviously, the generalized Heisenberg equation is a revisionary theory associated with the structure function $s$ of the known Heisenberg equation, it means that the classical Heisenberg equation is incomplete for quantum mechanics, because of the missing term $-\frac{1}{\sqrt{-1}\hbar }\hat{H}{{\left[ s,\hat{f} \right]}_{QPB}}$. On the other hand, the G-dynamics is completely induced by the covariant dynamics as an independent dynamics. The equilibrium equation of the generalized Heisenberg equation is given by $\frac{d\hat{f}}{dt}=0$, in other words,
\[{{\left[ \hat{f},\hat{H} \right]}_{QPB}}=\hat{H}{{\left[ s,\hat{f} \right]}_{QPB}}\]In the same way, the equilibrium equation of the covariant dynamics is $\frac{\mathcal{D}\hat{f}}{dt}=0$, that is, the covariant equilibrium equation of the covariant dynamics is given by $\left[ \hat{f},\hat{H} \right]=0$, it leads to a quantum covariant conservative quantity, fundamentally.

\subsection{Canonical quantum covariant Hamilton equations}
For the sake of argument that we manage to consider the position operator and momentum operator respectively.

The generalized Heisenberg equations and canonical quantum covariant Hamilton equations of motion on manifolds can be written respectively
  \begin{description}
    \item[the generalized Heisenberg equations of motion]
    $$d{{x}_{j}}/dt=\frac{1}{\sqrt{-1}\hbar }{{\left[ {{x}_{j}},\hat{H} \right]}_{QPB}}\begin{matrix}
   , & {}  \\
\end{matrix}d{{\hat{p}}_{j}}/dt=\frac{1}{\sqrt{-1}\hbar }\left( {{\left[ {{{\hat{p}}}_{j}},\hat{H} \right]}_{QPB}}-\hat{H}{{\left[ s,{{{\hat{p}}}_{j}} \right]}_{QPB}} \right)$$

    \item[canonical quantum covariant Hamilton equations]
  $$\mathcal{D}{{x}_{j}}/dt=\frac{1}{\sqrt{-1}\hbar }\left[ {{x}_{j}},\hat{H} \right],\begin{matrix}
   {} & \mathcal{D}{{{\hat{p}}}_{j}}/dt=\frac{1}{\sqrt{-1}\hbar }\left[ {{{\hat{p}}}_{j}},\hat{H} \right]  \\
\end{matrix}$$

  \end{description}
where $\left[ \cdot ,~\cdot  \right]={{\left[ \cdot ,~\cdot  \right]}_{QPB}}+G\left( s,~\cdot ,~\cdot  \right)$ is quantum covariant Poisson bracket.

When the system of the spring is very small then we have to use a quantum mechanical description of the system to correctly take into account quantum effects such as the quantization of energy levels. We will follow the procedure of canonical covariant quantization to obtain such a general description: namely, we impose the equal-time geometric canonical commutation relation.

\subsection{The QCPB for quantum harmonic oscillator}
As a certainly example based on the theorem \ref{t2}, we will now start by briefly reviewing the quantum mechanics of a one-dimensional harmonic
oscillator \eqref{eq11}, and see how the QCPB can be incorporated using GGC in the covariant quantization procedure.
With the Hamiltonian given by \eqref{eq11}, let's use the QCPB to recalculate the \eqref{eq10}, we can see how difference emerges. More specifically, the covariant dynamics in terms of the position reads
\[\frac{\mathcal{D}}{dt}x\left( t \right)=\frac{\sqrt{-1}}{\hbar }\left[ \hat{H}^{\left( cl \right)},x\left( t \right) \right]=\frac{{{\hat{p}}^{\left( cl \right)}}\left( t \right)}{m}+x\left( t \right){{\hat{w}}^{\left( cl \right)}}\]
By direct computation, the G-dynamics in terms of $\hat{H}^{\left( cl \right)}$ is given by
\begin{align}\label{eq6}
 {{\hat{w}}^{\left( cl \right)}}&=-\sqrt{-1}{{\left[ s,{{\hat{H}}^{\left( cl\right)}}\right]}_{QPB}}/\hbar=\frac{\sqrt{-1}}{\hbar }\frac{{{\hat{p}}^{\left( cl \right)2}}s+2{{\hat{p}}^{\left( cl \right)}}s{{\hat{p}}^{\left( cl \right)}}}{2m}\\
 &=b_{c}\left( 2u\frac{d}{dx}+u_{x} \right)\notag
\end{align}where $b_{c}=-\frac{\sqrt{-1}\hbar }{2m}$, and $u=\frac{ds}{dx}$, $u_{x}=\frac{{{d}^{2}}}{d{{x}^{2}}}s$ are used. Note that the G-dynamics in terms of $\hat{H}^{\left( cl \right)}$ describes the quantum rotation of the manifolds space in a certainty quantum system, its eigenvalues represent the frequency spectrum.
Furthermore, it gets $\sqrt{-1}\hbar {{{\hat{w}}}^{\left( cl \right)}}={{\left[ s,{{\hat{H}}^{\left( cl\right)}}\right]}_{QPB}}$, then it leads to
$$i\hbar {{\hat{w}}^{\left( cl \right)}}\psi=\frac{{{\hbar }^{2}}}{2m}\left( 2u{{\psi }_{x}}+\psi {{u}_{x}} \right)$$And the generalized Heisenberg equation follows
$\frac{d}{dt}x\left( t \right)=\frac{{{\hat{p}}^{\left( cl \right)}}\left( t \right)}{m}$.
And the generalized Heisenberg equation with respect to $x$ follows
\begin{align}
 \frac{d}{dt}x\left( t \right) & =\frac{\sqrt{-1}}{\hbar }{{\left[ \hat{H}^{\left( cl \right)},x\left( t \right) \right]}_{QPB}}+\frac{\sqrt{-1}}{\hbar }\hat{H}^{\left( cl \right)}{{\left[ s,x\left( t \right) \right]}_{QPB}} \notag\\
 & =\frac{{{\hat{p}}^{\left( cl \right)}}\left( t \right)}{m}\notag
\end{align}where ${{\left[ s,x\left( t \right) \right]}_{QPB}}=0$.
In the same way,  the covariant dynamics for the classical momentum operator is
\begin{align}
  \frac{\mathcal{D}}{dt}{{{\hat{p}}}^{\left( cl \right)}}\left( t \right)&=\frac{\sqrt{-1}}{\hbar }\left[\hat{H}^{\left( cl \right)},{{{\hat{p}}}^{\left( cl \right)}}\left( t \right) \right] \notag\\
 & =-m{{\omega }^{2}}x-\hat{H}^{\left( cl \right)}u+{{\hat{p}}^{\left( cl \right)}}\left( t \right){{\hat{w}}^{\left( cl \right)}} \notag
\end{align}
Accordingly, the generalized Heisenberg equation in terms of the ${{\hat{p}}}^{{\left( cl \right)}}$ appears
\begin{align}
 \frac{d}{dt}{{{\hat{p}}}^{\left( cl \right)}}\left( t \right) &=\frac{\sqrt{-1}}{\hbar }{{\left[ \hat{H}^{\left( cl \right)},{{{\hat{p}}}^{\left( cl \right)}}\left( t \right) \right]}_{QPB}}+\frac{\sqrt{-1}}{\hbar }\hat{H}^{\left( cl \right)}{{\left[ s,{{{\hat{p}}}^{\left( cl \right)}}\left( t \right) \right]}_{QPB}} \notag\\
 & =-m{{\omega }^{2}}x-\hat{H}^{\left( cl \right)}u \notag
\end{align}
where $\sqrt{-1}\hbar u={{\left[ s,{{{\hat{p}}}^{\left( cl \right)}}\left( t \right) \right]}_{QPB}}$.

\subsection{The GSPB and the QCPB}
In total, we emphasize that the GSPB in an abstract form is given as definition shown
$\left\{ f,g \right\}={{\left\{ f,g \right\}}_{GPB}}+G\left(s, f,g \right)$,
While the QCPB is taken the same pattern to reflect the GSPB
$\left[ \hat{f},\hat{g} \right]={{\left[ \hat{f},\hat{g} \right]}_{QPB}}+G\left(s, \hat{f},\hat{g} \right)$ as a similar pattern for classical and quantum brackets.
In the view of the relation between quantum mechanics and classic mechanics. Notice that in this representation the fundamental equations are similar to the classical equations with the substitution.
\begin{align}
  & \left\{ f,g \right\}={{\left\{ f,g \right\}}_{GPB}}+G\left( s,f,g \right) \notag\\
 & \begin{matrix}
   {} & {} & {}  & {} & {} \\
\end{matrix}\mapsto \frac{1}{\sqrt{-1}\hbar }\left[ \hat{f},\hat{g} \right]=\frac{1}{\sqrt{-1}\hbar }\left( {{\left[ \hat{f},\hat{g} \right]}_{QPB}}+G\left( s,\hat{f},\hat{g} \right) \right) \notag
\end{align}With the solid foundation of the GSPB and the QCPB in abstract, then the covariant dynamical equation certainly follows based on the GSPB and the QCPB respectively.
From the GCHS
\[\frac{\mathcal{D}f}{dt}={{\left\{ f,H \right\}}_{GPB}}+G\left(s, f,H \right)\]to the quantum covariant dynamics. Furthermore, the covariant equation of the motion of an arbitrary operator $\hat{f}$ or the observables $\hat{f}$ that satisfies the quantum covariant evolution
\[\frac{\mathcal{D}\hat{f}}{dt}=\frac{1}{\sqrt{-1}\hbar }\left( {{\left[ \hat{f},\hat{H} \right]}_{QPB}}+G\left(s, \hat{f},\hat{H} \right) \right)\]Namely, the time evolution of the operator $\hat{f}$ is then given by the corresponding equation
\[\frac{\mathcal{D}f}{dt}=\left\{ f,H \right\}\to \frac{\mathcal{D}\hat{f}}{dt}=\frac{1}{\sqrt{-1}\hbar }\left[ \hat{f},\hat{H} \right]\]
As a consequence, the equations of motion are respectively obtained by applying generalized Heisenberg equations of motion and canonical quantum covariant Hamilton equations.

According to the rule of quantization, we have the GSPB to QCPB given by
$\left\{ \cdot, \cdot\right\}\to \left[ \cdot, \cdot\right]$. More precisely,
$\left\{ \cdot, \cdot \right\}={{\left\{ \cdot, \cdot\right\}}_{GPB}}+G\left( s, \cdot, \cdot \right)$ is transformed to
\[\frac{1}{\sqrt{-1}\hbar }\left[  \cdot, \cdot \right]=\frac{1}{\sqrt{-1}\hbar }{{\left[  \cdot, \cdot \right]}_{QPB}}+\frac{1}{\sqrt{-1}\hbar }G\left(s, \cdot, \cdot \right)\]
More generally, this technique leads to a complete covariant quantization that can be accurately computed and analyzed, mathematically.

As a consequence, we put the geobracket and quantum geobracket together to contrastively analyze their connections.
Especially, the covariant correction term--the quantum geobracket $G\left( s, \hat{f},\hat{g} \right)$ is unknown, but by logically derivation, the specific formula of the geobracket $G\left(s, f,g \right)$ is surely concretized in GSPB, what we do is to copy this mode for the quantum geobracket $G\left(s, \hat{f},\hat{g} \right)$.
As described previously, the geobracket $G\left(s, f,g \right)$ in GSPB is defined as
 \[G\left(s,f,g \right) =f{{\left\{ s ,g \right\}}_{GPB}}-g{{\left\{ s ,f \right\}}_{GPB}}\]
In a similar mode for quantum mechanics, quantum geobracket $G\left( s, \hat{f},\hat{g} \right)$ can be basically concretized in QCPB, that is given by as defined
$$G\left(s,\hat{f},\hat{g} \right)=G\left(\hat{s},\hat{f},\hat{g} \right)=\hat{f}{{\left[ s,\hat{g} \right]}_{QPB}}-\hat{g}{{\left[ s,\hat{f} \right]}_{QPB}}$$

With the help of the structural function, the QCPB is well defined for covariant quantum mechanics\footnote{Notes: GCC: Geometric canonical commutation; CCHE:canonical covariant Hamilton equations; CD:covariant dynamics;   GHE: generalized Heisenberg equations; CTHE: Canonical thorough Hamilton equations.}.
\begin{center}
  \begin{tabular}{c r @{.} l}
\hline\hline
covariant quantum mechanics  \\
\hline
QCPB: $\left[ \hat{f},\hat{g} \right]={{\left[ \hat{f},\hat{g} \right]}_{QPB}}+G\left(s,\hat{f},\hat{g} \right)$, $G\left(s,\hat{f},\hat{g} \right)=\hat{f}{{\left[ s,\hat{g} \right]}_{QPB}}-\hat{g}{{\left[ s,\hat{f} \right]}_{QPB}}$ \\
\hline
GCC: $\left[ x_{i},{{{\hat{p}}}_{j }}\right]=\sqrt{-1}\hbar{{D}_{j}}{{x}_{i}}$, ${{D}_{i}}={{\partial }_{i}}+{{\partial }_{i}}s,{{\partial }_{i}}=\frac{\partial }{\partial {{x}_{i}}}$. \\
\hline
 Hamiltonian : $\hat{H}$  \\
\hline
CD:  $\frac{\mathcal{D}}{dt}{{x}_{i }}=\frac{1}{\sqrt{-1}\hbar }\left[ {{{x}}_{i}},\hat{H} \right],~~~\frac{\mathcal{D}}{dt}{{\hat{p}}_{j }}=\frac{1}{\sqrt{-1}\hbar }\left[ {{{\hat{p}}}_{j }},\hat{H} \right]$
\\
\hline
GHE: ${\dot{{x}_{i}}}=\frac{1}{\sqrt{-1}\hbar }{{\left[ {{{x}}_{i }},\hat{H} \right]}_{QPB}},~~\overset{\cdot }{\mathop{{{{\hat{p}}}_{j}}}}\, =\frac{1}{\sqrt{-1}\hbar }\left( {{\left[ {{{\hat{p}}}_{j }},\hat{H} \right]}_{QPB}}-\hat{H}\left[ s,{{{\hat{p}}}_{j }} \right]_{QPB} \right)$
\\
\hline
G-dynamics: $\hat{w}=\frac{1}{\sqrt{-1}\hbar }{{\left[ s,\hat{H} \right]}_{QPB}}$.\\
\hline\hline
\end{tabular}
\end{center}
where $\left[ \hat{f},\hat{g} \right]_{QPB}=\hat{f}\hat{g}-\hat{g}\hat{f}$ is quantum Poisson brackets for operators.

\section{Covariant quantization of Field}
In this section, we only depend on the QCPB to obtain the covariant quantization of field, generally, it helps us better understand how specifically the fields move and evolve.

For the quantum of a Bose field, the canonical coordinate ${{\varphi }_{i}}$ and the canonical conjugate momentum ${{p}_{i}}$ of the field satisfies the QCPB, it's depicted as
\[\left[ {{\varphi }_{i}},{{p}_{j}} \right]={{\left[ {{\varphi }_{i}},{{p}_{j}} \right]}_{QPB}}+G\left(s, {{\varphi }_{i}},{{p}_{j}} \right)\]
Subsequently,  the QCHS for Bose field follows
\[\left[ \varphi ,\hat{H} \right]={{\left[ \varphi ,\hat{H} \right]}_{QPB}}+G\left( s,\varphi ,\hat{H} \right)\]
Then canonical quantum covariant Hamilton equations emerge accordingly,
\[\frac{\mathcal{D}\varphi }{dt}=\frac{1}{\sqrt{-1}\hbar }\left( {{\left[ \varphi ,\hat{H} \right]}_{QPB}}+G\left(s, \varphi ,\hat{H} \right) \right)\]
Similarly, meanwhile,  for the quantum of Fermi field $\Psi \left( t,x \right)$, and the canonical coordinates and canonical conjugate momenta of the field are $\Psi \left( t,x \right)$, $\pi \left( t,x \right)$ respectively.
The same procedure as Bose field does goes for Fermi field, we list it as follows,  the QCHS for Fermi field $\Psi \left( t,x \right)$ is given by
$\left[ \Psi ,\hat{H} \right]={{\left[ \Psi ,\hat{H} \right]}_{QPB}}+G\left(s, \Psi ,\hat{H} \right)$ and quantum covariant Hamilton equation appears
\[\frac{\mathcal{D}\Psi }{dt}=\frac{1}{\sqrt{-1}\hbar }\left( {{\left[ \Psi ,\hat{H} \right]}_{QPB}}+G\left(s, \Psi ,\hat{H} \right) \right)\]
\[\frac{\mathcal{D}\pi }{dt}=\frac{1}{\sqrt{-1}\hbar }\left( {{\left[ \pi ,\hat{H} \right]}_{QPB}}+G\left(s, \pi ,\hat{H} \right) \right)\]
In general, the quantum geobracket are respectively given by
$G\left(s, \varphi ,\hat{H} \right)\equiv \varphi {{\left[ s,\hat{H} \right]}_{QPB}}-\hat{H}{{\left[ s,\varphi  \right]}_{QPB}}$ and
$G\left( s,\Psi ,\hat{H} \right)\equiv \Psi {{\left[ s,\hat{H} \right]}_{QPB}}-\hat{H}{{\left[ s,\Psi  \right]}_{QPB}}$.
This procedure can be applied to the covariant quantization of any field theory: whether of fermions or bosons.  All other fields can be quantized by QCPB of this procedure.

\subsection{Canonical covariant quantization of Bose field}
The quantum of a Bose field, the spin is integer or zero, obeying the Bose-Einstein statistical method, known as bosons, the canonical coordinate ${{\varphi }_{i}}$ and the canonical conjugate momentum ${{p}_{i}}$ of the field are regarded as the operator of Hilbert space, and the corresponding geometric canonical commutation relation in discrete field form is
$\left[ {{\varphi }_{i}},{{p}_{j}} \right]=\sqrt{-1}\hbar {{D}_{j}}{{\varphi}_{i}}$.
Accordingly, the generalized canonical equations of motion for corresponding quantum fields are shown as
$$\overset{\centerdot }{\mathop{\varphi }}\,=\frac{\sqrt{-1}}{\hbar }\left[ \hat{H},\varphi  \right]_{QPB}+\frac{\sqrt{-1}}{\hbar } \hat{H}\left[s,\varphi  \right]_{QPB}$$$$\overset{\centerdot }{\mathop{\pi }}\,=\frac{\sqrt{-1}}{\hbar }\left[ \hat{H},\pi  \right]_{QPB}+\frac{\sqrt{-1}}{\hbar } \hat{H}\left[s,\pi  \right]_{QPB}$$The quantization of the field depends entirely on the commutators.
And canonical quantum covariant Hamilton equations are accordingly shown as
\[\frac{\mathcal{D}}{dt}\varphi =\frac{\sqrt{-1}}{\hbar }\left[ \hat{H},\varphi  \right],~~\frac{\mathcal{D}}{dt}\pi =\frac{\sqrt{-1}}{\hbar }\left[ \hat{H},\pi  \right]\]

\subsection{Canonical covariant quantization of Fermi field}
The quantum of Fermi field has half spin and obeys the Fermi-Dirac statistical method. Suppose that $\Psi \left( t,x \right)$ is the field function of the Fermi field, and the canonical coordinates and canonical conjugate momenta of the field are $\Psi \left( t,x \right)$, $\pi \left( t,x \right)$ respectively.
Accordingly, the generalized canonical equations of motion for corresponding quantum fields are shown as
$$\overset{\centerdot }{\mathop{\Psi }}\,=\frac{\sqrt{-1}}{\hbar }\left[ \hat{H},\Psi  \right]_{QPB}+\frac{\sqrt{-1}}{\hbar } \hat{H}\left[s,\Psi  \right]_{QPB}$$$$\overset{\centerdot }{\mathop{\pi }}\,=\frac{\sqrt{-1}}{\hbar }\left[ \hat{H},\pi  \right]_{QPB}+\frac{\sqrt{-1}}{\hbar } \hat{H}\left[s,\pi  \right]_{QPB}$$
The canonical covariant motion equation of quantum field is
\begin{align}
  & \frac{\mathcal{D}}{dt}\Psi \left( t,x \right)=\frac{\sqrt{-1}}{\hbar }\left[ \hat{H}\left( t \right),\Psi \left( t,x \right) \right]=\overset{\centerdot }{\mathop{\Psi }}\,+\Psi \hat{w} \notag\\
 & \frac{\mathcal{D}}{dt}\pi \left( t,x \right)=\frac{\sqrt{-1}}{\hbar }\left[ \hat{H}\left( t \right),\pi \left( t,x \right) \right]=\overset{\centerdot }{\mathop{\pi }}\,+\pi \hat{w}\notag
\end{align}

\section{Geomentum operator and geometric canonical commutation}
\begin{definition}\label{d3}
  Let $M$ be a smooth manifold represented by geometric potential function $s$, then geomentum operator is defined as
  $\hat{p}=-\sqrt{-1}\hbar D$, where $D=\nabla +\nabla s$. The component is ${{\hat{p}}_{j}}=-\sqrt{-1}\hbar {{D}_{j}}$, where ${{D}_{j}}={{\partial }_{j}}+{{\partial }_{j}}s$ holds, ${{\partial }_{j}}=\frac{\partial }{\partial {{x}_{j}}}$.
\end{definition}
For instance, let's consider the canonical commutation relation Eq\eqref{eq1}, namely, let $\hat{f}={{{x}_{i}}},~\hat{g}=\hat{{{p}_{j}}}$ be given for discussions. As a result, geometric canonical commutation relation can be compactly written as
\[\left[ {{{x}_{i}}},\hat{{{p}_{j}}} \right]= \sqrt{-1}\hbar{{D}_{j}}{{x}_{i}}\]
Let $\hat{f}={{x}_{i}},~\hat{g}={{\hat{p}}_{j}}=-\sqrt{-1}\hbar {{D}_{j}}$, by using QCPB, we have \[\left[ {{x}_{i}},-\sqrt{-1}\hbar {{D}_{j}} \right]={{\left[ {{x}_{i}},-\sqrt{-1}\hbar {{D}_{j}} \right]}_{QPB}}+G(s,{{x}_{i}},-\sqrt{-1}\hbar {{D}_{j}})\]More precisely, it shows
\[\left[ {{x}_{i}},-\sqrt{-1}\hbar {{D}_{j}} \right]={{\left[ {{x}_{i}},-\sqrt{-1}\hbar {{D}_{j}} \right]}_{QPB}}+{{x}_{i}}{{\left[ s,-\sqrt{-1}\hbar {{D}_{j}} \right]}_{QPB}}\]where we have used ${{\left[ s,{{{x}_{i}}} \right]}_{QPB}}=0$.
Further, put the wave function onto it, it leads to the result
\begin{align}
 \left[ {{x}_{i}},-\sqrt{-1}\hbar {{D}_{j}} \right]\psi &=-\sqrt{-1}\hbar {{\left[ {{x}_{i}},{{D}_{j}} \right]}_{QPB}}\psi -\sqrt{-1}\hbar {{x}_{i}}{{\left[ s,{{D}_{j}} \right]}_{QPB}}\psi \notag \\
 & =-\sqrt{-1}\hbar \left\{ {{\left[ {{x}_{i}},{{D}_{j}} \right]}_{QPB}}\psi +{{x}_{i}}{{\left[ s,{{D}_{j}} \right]}_{QPB}}\psi  \right\} \notag\\
 & =\sqrt{-1}\hbar \left( {{\delta }_{ij}}+{{x}_{i}}{{\partial }_{j}}s \right)\psi \notag \\
 & =\sqrt{-1}\hbar \psi {{D}_{j}}{{x}_{i}} \notag
\end{align}
As a result, we obtain
\[\left[ {{x}_{i}},{{\hat{p}}_{j}} \right]=\left[ {{x}_{i}},-\sqrt{-1}\hbar {{D}_{j}} \right]=\sqrt{-1}\hbar {{D}_{j}}{{x}_{i}}\]

Equivalently, as defined above, using QCPB with ${{\hat{{{p}_{j}}}}^{\left( cl \right)}}=-\sqrt{-1}\hbar \frac{\partial }{\partial {{x}_{j}}},~~\hat{{{x}_{i}}}\equiv {{x}_{i}}$, then the QCPB is
\[\left[ {{{x}_{i}}},{{\hat{{{p}_{j}}}}^{\left( cl \right)}} \right]={{\left[ {{{x}_{i}}},{{\hat{{{p}_{j}}}}^{\left( cl \right)}} \right]}_{QPB}}+{{{x}_{i}}}{{\left[ s,{{\hat{{{p}_{j}}}}^{\left( cl \right)}} \right]}_{QPB}}-{{\hat{{{p}_{j}}}}^{\left( cl \right)}}{{\left[ s,{{{x}_{i}}} \right]}_{QPB}}\]
One only needs to calculate the quantum geobracket
\[G\left(s,{{{x}_{i}}},{{\hat{{{p}_{j}}}}^{\left( cl \right)}} \right)\psi ={{{x}_{i}}}{{\left[ s,{{\hat{{{p}_{j}}}}^{\left( cl \right)}} \right]}_{QPB}}\psi -{{\hat{{{p}_{j}}}}^{\left( cl \right)}}{{\left[ s,{{{x}_{i}}} \right]}_{QPB}}\psi \]
Due to ${{\left[ s,{{{x}_{i}}} \right]}_{QPB}}=0$ is obvious, then
\begin{align}
 {{{x}_{i}}}{{\left[ s,{{\hat{{{p}_{j}}}}^{\left( cl \right)}} \right]}_{QPB}}\psi &={{x}_{i}}s{{\hat{{{p}_{j}}}}^{\left( cl \right)}}\psi -{{x}_{i}}{{\hat{{{p}_{j}}}}^{\left( cl \right)}}\left( s\psi  \right) \notag\\
 & =-{{x}_{i}}s\sqrt{-1}\hbar \frac{\partial \psi }{\partial {{x}_{j}}}+{{x}_{i}}\sqrt{-1}\hbar \frac{\partial }{\partial {{x}_{j}}}\left( s\psi  \right) \notag\\
 & ={{x}_{i}}\psi \sqrt{-1}\hbar \frac{\partial }{\partial {{x}_{j}}}s \notag
\end{align}
Thus, it leads to the result \[G\left( s, {{{x}_{i}}},{{\hat{{{p}_{j}}}}^{\left( cl \right)}}\right)\psi ={{x}_{i}}\psi \sqrt{-1}\hbar \frac{\partial }{\partial {{x}_{j}}}s\]which also deduces
$G\left(s,{{{x}_{i}}},{{\hat{{{p}_{j}}}}^{\left( cl \right)}} \right)=\sqrt{-1}\hbar {{x}_{i}}\frac{\partial }{\partial {{x}_{j}}}s$. As a result, one can conclude that
\begin{align}
 \left[ {{{x}_{i}}},{{\hat{{{p}_{j}}}}^{\left( cl \right)}} \right] & ={{\left[ {{{x}_{i}}},{{\hat{{{p}_{j}}}}^{\left( cl \right)}} \right]}_{QPB}}+{{{x}_{i}}}{{\left[ s,{{\hat{{{p}_{j}}}}^{\left( cl \right)}} \right]}_{QPB}}
=\sqrt{-1}\hbar {{\delta }_{ij}}+G\left( s, {{{x}_{i}}},{{\hat{{{p}_{j}}}}^{\left( cl \right)}}\right) \notag\\
 & =\sqrt{-1}\hbar \left( {{\delta }_{ij}}+{{x}_{i}}\frac{\partial }{\partial {{x}_{j}}}s \right) \notag
\end{align}
The situation for this same result is because ${{\left[ s,{{u}_{j}} \right]}_{QPB}}=0$ holds, where ${{u}_{j}}=\frac{\partial }{\partial {{x}_{j}}}s $.

 Geometric canonical commutation relation can be called geometric canonical quantization rules. The
  geometric equal-time canonical commutation relation is
  \[\left[ {{{x}_{i}}},\hat{{{p}_{j}}} \right]=\sqrt{-1}\hbar {{D}_{j}}{{x}_{i}}\]
Geometric canonical commutation relation can be expressed in a specific form \[\left[ {{{x}_{i}}},\hat{{{p}_{j}}} \right]= \sqrt{-1}\hbar{{D}_{j}}{{x}_{i}}=\sqrt{-1}\hbar \left( {{\delta }_{ij}}+{{x}_{i}}\frac{\partial }{\partial {{x}_{j}}}s \right) \]
In other words, it also can be rewritten as
$\left[ {{x}_{i}},\hat{{{p}_{j}}} \right]=\sqrt{-1}\hbar {{\theta }_{ij}}$, where
${{\theta}_{ij}}={{\delta }_{ij}}+{{x}_{i}}{{\partial }_{j}}s$, and ${{\partial }_{j}}=\frac{\partial }{\partial {{x}_{j}}}$.
Accordingly, the quantum geobracket relates to the below
$\frac{1}{\sqrt{-1}\hbar }G\left( s,{{x}_{i}},{{\hat{{{p}_{j}}}}} \right)={{\theta }_{ij}}-{{\delta }_{ij}}$.

\section{Conclusions}

The conclusions can be assembled here. In this paper, by bringing the structure function determined by the spacetime itself into the commutator,  then a generalized geometric commutator totally relied on the form of geometric potential function, in other words, generalized geometric commutator is constructed by the commutator equipped with geomutator. Accordingly, quantum covariant Poisson bracket and quantum geometric bracket are respectively defined by generalized geometric commutator and geomutator.

In theory, with the definition of quantum covariant Poisson bracket (QCPB) and generalized structure Possion bracket(GSPB), it well defines the covariant theoretical framework for the quantum mechanics and analytical mechanics.

Firstly, by using the geometric potential function $s$ determined by the manifolds, one generalized the QPB to more general form QCPB or GGC, it's formally shown as
\[\left[ \cdot ,~\cdot  \right]={{\left[ \cdot ,~\cdot  \right]}_{QPB}}+G\left(s, \cdot ,~\cdot  \right)\]
where $G\left(s,\cdot ,~\cdot  \right)=\cdot {{\left[ s,~\cdot  \right]}_{QPB}}+\cdot {{\left[ \cdot~ ,s \right]}_{QPB}}$ is quantum geobracket as a  correctional term for the QPB, which is the vital part of the QCPB to the quantum mechanics. Geometric canonical commutation relation follows as well.  Secondly, using QCPB to define QCHS, and the covariant dynamics, the generalized Heisenberg equation, G-dynamics are emerged one by one. In the end, the quantum covariant Poisson bracket (QCPB) is used for  quantization of field including Bose field and Fermi field, we obtain
covariant quantization of field, we revise the classical theory for both
Bose field and Fermi field. As for the application of quantum covariant Poisson bracket, the classical canonical commutation relation and quantization of field are considered under the QCPB, we obtain some different useful results.


\end{document}